\documentclass[reprint,preprintnumbers,amsmath,amssymb,aps,nofootinbib,superscriptaddress]{revtex4-2}

\usepackage[utf8]{inputenc}  
\usepackage{graphicx}  
\usepackage{xcolor}  
\usepackage{amsmath}  
\usepackage{amsfonts}  
\usepackage{latexsym}  
\usepackage{amssymb,bbm}  
\usepackage{bm}  
\usepackage[colorlinks=true,allcolors=blue]{hyperref}  
\usepackage{setspace}
\usepackage{braket}
\usepackage{float}

\begin{document}

\title{Continuous-Time Random Walk Description of Anomalous Spin Transport in Dilute Dipolar Networks}

\author{Cooper M. Selco}
\affiliation{Department of Chemistry, University of California, Berkeley, Berkeley, CA 94720, USA}

\author{Christian Bengs}
\affiliation{School of Chemistry, University of Southampton, Southampton, SO17 1BJ, UK}

\author{Ashok Ajoy}
\email{ashokaj@berkeley.edu}
\affiliation{Department of Chemistry, University of California, Berkeley, Berkeley, CA 94720, USA}
\affiliation{Chemical Sciences Division, Lawrence Berkeley National Laboratory, Berkeley, Berkeley, CA 94720, USA}

\begin{abstract}
Nuclear spin diffusion is often summarized by a single diffusion coefficient, but this coarse-grained description can fail in dilute solids where positional disorder and long-range dipolar couplings generate a broad distribution of hopping rates. We develop a continuous-time random-walk (CTRW) description of $^{13}$C polarization transport in natural-abundance diamond (1.1\%), constructing the rate matrix from dipolar-mediated flip-flop couplings and sampling exact continuous-time trajectories. Although site-to-site hopping is Markovian, the disorder-averaged dynamics give rise to emergent, anomalous transport. The empirical waiting-time distribution exhibits a heavy tail with exponent $\alpha=0.64$ and exponential cutoff $t_{\rm cutoff}=19$ s; the mean jump length becomes correlated with the waiting time $\tau$ for $\tau\gtrsim0.1$ s; and the mean-squared displacement grows sublinearly in both step number and physical time, with exponents $\gamma=0.56$ and $\delta=0.87$ respectively. We trace the microscopic origin of these signatures to geometric trapping: polarization can rapidly exchange within strongly coupled clusters, including dimers, while weak inter-cluster links control long-range exploration. A kinetic percolation construction links global transport to inter-cluster crossing times, and identifies a corresponding crossing time of $\sim20$ s, consistent with $t_{\rm cutoff}$. Finally, mapping paramagnetic impurities onto hard-sphere traps connects the CTRW framework to classic studies of trapping in reaction-diffusion theory and reproduces the qualitative timescale of experimentally measured relaxation, whereas a continuum diffusion equation description does not. These results show that dilute dipolar spin networks require a microscopic, network-resolved transport description beyond the Fickian diffusion equation.
\end{abstract}

\maketitle


\section{Introduction}
Transport in disordered media is a recurring theme across chemical physics, condensed matter physics, and magnetic resonance~\cite{kirkpatrickClassicalTransportDisordered1971, sahimiStochasticTransportDisordered1983, watsonCharacterizingPorousMedia1997, ben-avrahamDiffusionReactionsFractals2000,  hoflingAnomalousTransportCrowded2013, huberSoftMatterHard2015}. In coupled nuclear spin systems, the transported quantity is spin polarization, which migrates through dipolar-mediated flip-flop processes between like nuclei. The standard macroscopic description of this process is the spin-diffusion equation, originally developed for dense crystalline spin networks and later applied broadly to relaxation and polarization transfer in solids~\cite{abragamPrinciplesNuclearMagnetism1983, bloembergenInteractionNuclearSpins1949c, blumbergNuclearSpinLatticeRelaxation1960c, simmonsNuclearSpinLatticeRelaxation1962a, furmanSpinDiffusionSpinlattice1999b, ramanathanDynamicNuclearPolarization2008b}. This description assumes that the microscopic hopping process can be coarse grained into a spatially homogeneous, time-independent transport coefficient. Under these conditions, polarization spreads with a Gaussian spatial profile and the mean squared displacement grows linearly in time~\cite{hughesRandomWalksRandom1995, bouchaudAnomalousDiffusionDisordered1990}.

These assumptions may become invalid in dilute, positionally disordered spin networks. For nuclear spins coupled by magnetic dipolar interactions, the flip-flop matrix element scales as $r^{-3}$ and the corresponding hopping rate scales approximately as $r^{-6}$ when the relevant spectral overlap is treated as fixed~\cite{abragamPrinciplesNuclearMagnetism1983, ernstSpinDiffusionSolids1998}. Random placement of a rare isotope therefore produces a broad distribution of local escape rates: some spins belong to tightly coupled clusters, while others are connected to the rest of the network only through weak, long-range links. In such a landscape, the relevant question is whether a time-independent diffusion coefficient is an appropriate coarse-grained variable at all. 

Natural-abundance diamond provides a particularly useful setting in which to explore this problem. It is a concrete realization of a broader class of sparse, long-range-coupled networks in which local structure and weak inter-cluster links jointly determine transport. The $^{13}$C nuclei form a dilute spin-$1/2$ network at 1.1\% natural abundance on the diamond lattice. Understanding polarization transport in this network is relevant for dynamic nuclear polarization (DNP)~\cite{ajoyOrientationindependentRoomTemperature2018c, paglieroOpticallyPumpedSpin2020, priscoScalingAnalysesHyperpolarization2021a, vonwitteNuclearSpinDiffusion2025}, nuclear relaxation~\cite{beatrezElectronInducedNanoscale2023d, selcoEmergentDecoherenceDynamics2025a, selcoBreakdownDisorderSuppressedFloquet2026, shahTunableMpembaEffect2026}, and the performance of diamond-based quantum sensing and memory platforms~\cite{sahinHighFieldMagnetometry2022d, moonSensingDiscreteTime2026, richterRobustQuantumSensing2026}.

To describe transport in such disordered systems, previous studies have primarily employed ensemble-level or master-equation descriptions, averaging over random spin configurations to obtain effective diffusion constants and transport behavior~\cite{vugmeisterSpatialSpectralSpin1976c,dzheparovCrossRelaxationDilute,goldmanNuclearSpinDiffusion1982b}. Subsequent work developed a range of asymptotic and approximate frameworks for spin transport in disordered systems, including propagator-based descriptions and systematic expansions aimed at capturing different time regimes~\cite{dzheparovTransportSpinPolarization1991, dzheparovRandomWalksDisordered1993,dzheparovDelocalizationExcitationsDisordered2005}. A recurring challenge in these approaches is that fully microscopic, disorder-averaged solutions are generally intractable, which has led to a focus on asymptotic limits or effective approximations. More recently, experimental works have reported non-Fickian signatures such as sub-diffusive scaling and non-Gaussian hydrodynamic profiles~\cite{selcoEmergentDecoherenceDynamics2025a, zuEmergentHydrodynamicsStrongly2021b}, highlighting the need for microscopic descriptions that directly connect network geometry to the observed transport laws.

Motivated by these challenges, microscopic simulation approaches have also been developed. Ref.~\cite{karabanovDynamicNuclearPolarization2015} derived an effective classical master equation in the Zeeman subspace from the full quantum dynamics and combined it with kinetic Monte Carlo simulations to study polarization transport in large spin systems~\cite{wisniewskiSolidEffectDNP2016}. Additionally, dynamic mean-field theory has been explored as a promising route toward first-principles descriptions of spin diffusion in interacting solids, providing a microscopic description of local polarization exchange while extensions to long-range transport remain an active area of development~\cite{grasserUnderstandingDynamicsRandomly2023, grasserFirstprinciplesSimulationSpin2026}

Here we develop a continuous-time random walk (CTRW) description of polarization transport in the dilute $^{13}$C network. For each realization of the positional disorder, we construct a dipolar hopping rate matrix $\mathbf{W}$ and sample exact stochastic trajectories using a Gillespie algorithm~\cite{gillespieGeneralMethodNumerically1976, gillespieExactStochasticSimulation1977}. This approach retains the microscopic lattice geometry while remaining directly connected to the usual diffusion-equation description in the homogeneous limit. In contrast to previous work, the present approach focuses on a trajectory-based formulation of transport, in which polarization dynamics are analyzed at the level of individual stochastic trajectories. In particular, we note that the kinetic Monte Carlo approach of Ref.~\cite{karabanovDynamicNuclearPolarization2015} also samples stochastic polarization trajectories. Their work is primarily focused on spin transport in the context of DNP, with transition rates that depend on the instantaneous spin configuration and can change in time. Importantly, the main focus in that work is on ensemble-averaged quantities such as the diffusion constant. In the present work, we consider a positionally disordered network and transition rates that are fixed for each random lattice configuration, which allows us to focus on the role of the quenched positional disorder. Importantly, the primary focus in this work is on microscopic transport statistics—including waiting times, jump lengths, spatiotemporal coupling, repeated site revisits, and geometric confinement—that are typically obscured in ensemble-level descriptions, but are important for understanding transport in strongly disordered networks.

We find that the resulting transport is strongly anomalous over experimentally relevant timescales. The waiting-time distribution develops a heavy tail with an exponential cutoff near 20 s, long waiting events become correlated with longer jumps, and the mean squared displacement grows sub-linearly both with the number of hops and with physical time. The microscopic origin is geometric trapping: polarization can undergo many rapid back-and-forth transitions within small, strongly coupled clusters before escaping through weak inter-cluster links. We support this picture using both trajectory statistics and fully quantum simulations of small disordered spin clusters, and show that the anomalous transport remains unchanged under a coarse-graining over strongly coupled clusters. To quantify the relevant connectivity and link global transport to cluster-resolved dynamics, we introduce a kinetic percolation framework. This analysis identifies a characteristic inter-cluster crossing time of $\sim20$ s, consistent with the cutoff in the waiting time distribution. Finally, we incorporate paramagnetic impurities as hard-sphere traps and show that the resulting lattice trapping dynamics reproduces the timescale of experimentally observed relaxation, whereas a continuum diffusion-equation model does not. Together, these results show that spin transport in dilute, positionally disordered networks is governed by microscopic network geometry and cannot, in general, be reduced to conventional Fickian diffusion.

\section{Microscopic model and stochastic framework}

We formulate nuclear polarization transport as a continuous-time Markov jump process on a fixed realization of the $^{13}$C lattice. This section establishes the connection between three descriptions used throughout the paper: the continuum diffusion equation, the site-resolved master equation, and the stochastic trajectories sampled in the simulations. The diffusion equation is recovered only after imposing homogeneity, decoupling of space and time, and finite transport moments. The dilute dipolar networks studied here retain the same local Markovian hopping rule, but can violate these coarse-graining assumptions after averaging over quenched positional disorder.

\subsection{Continuous-time random-walk formulation}

A continuous-time random walk (CTRW) is specified by a sequence of spatial jumps and the waiting times separating them. We denote a jump vector by $\mathbf{r}$, its length by $\ell=|\mathbf{r}|$, and the waiting time before the next jump by $\tau$. In a homogeneous CTRW, these variables are drawn from a jump distribution $\phi(\mathbf{r})$ and a waiting-time distribution $\psi(\tau)$. More generally, the process is described by a joint distribution $\chi(\mathbf{r},\tau)$. If $\chi(\mathbf{r},\tau)=\phi(\mathbf{r})\psi(\tau)$, spatial and temporal increments are statistically independent and the walk is decoupled. If this factorization fails, long waiting events may be correlated with particular jump lengths or directions, producing a coupled CTRW~\cite{klafterFirstStepsRandom2011a}.


To apply this framework to spin transport, we represent the dynamics as an ensemble of ``random walkers'' on the nuclear spin lattice. In this picture, polarization can be thought of as being decomposed into localized packets, each represented by a random walker undergoing stochastic hops in time between nuclear spins. Waiting times between jumps correspond to intervals between dipolar flip-flop events, while jump lengths correspond to the distances between spins that exchange polarization


To lay the background for what follows, we first show how the standard diffusion equation follows from the CTRW under suitable assumptions. The ensemble-averaged dynamics generated by the stochastic trajectories of the CTRW can be equivalently described in terms of a master equation for the site-resolved polarization. In this description, we focus on the quantity $P(\mathbf{r},t)$, which is the probability of finding a walker at site $\mathbf{r}$ at time $t$. Under the assumption of a homogeneous, decoupled CTRW—where the waiting-time and jump-length distributions are independent and identical across sites—the evolution of $P(\mathbf{r},t)$ can be written as,
\begin{equation}
\begin{aligned}
P(\mathbf{r},t) &= P(\mathbf{r},0)\Psi(t)
+ \int_{0}^{t} dt'\,\psi(t-t')\sum_{\mathbf{r}'} \phi(\boldsymbol{\mathbf{r}-\mathbf{r}'})P(\mathbf{r}',t'),
\label{GME}
\end{aligned}
\end{equation}
where
\begin{equation}
\begin{aligned}
\Psi(t)=1-\int_{0}^{t}\psi(\tau)d\tau
\label{eq:survival_function}
\end{aligned}
\end{equation}
is the probability that no jump has occurred during an interval of length $t$. In Eq.~\ref{GME}, the first term accounts for walkers that have not yet moved from site $\mathbf{r}$, while the second term describes arrivals at $\mathbf{r}$ from all
other sites $\mathbf{r}'$ after a jump occurring at time $t'$, followed by a waiting time $t-t'$.

Assuming Markovian hopping dynamics, the process is memoryless, and the escape from each site is described by a Poisson process with constant rate. Consequently, the waiting-time distribution takes the exponential form,
\begin{equation}
\begin{aligned}
\psi(\tau)=\lambda e^{-\lambda\tau},
\label{eq:poisson_waiting}
\end{aligned}
\end{equation}
where $\lambda=\langle \tau \rangle ^{-1}$ is the average escape rate. Substituting this form into Eq.~\ref{GME} and differentiating with respect to time gives
\begin{equation}
\begin{aligned}
\frac{\partial P(\mathbf{r},t)}{\partial t}= -\lambda P(\mathbf{r},t) + \lambda\sum_{\mathbf{r}'}\phi(\mathbf{r}-\mathbf{r}')P(\mathbf{r}',t).
\label{differentiate_gme}
\end{aligned}
\end{equation}
To obtain the continuum limit, the lattice sum can be replaced by an integral, and $P(\mathbf{r}',t)$ can be expanded about $\mathbf{r}$ as~\cite{kampenStochasticProcessesPhysics2011}
\begin{equation}
\begin{aligned}
P(\mathbf{r}',t) \approx P(\mathbf{r},t) + \nabla P(\mathbf{r},t)(\mathbf{r}'-\mathbf{r}) + \frac{1}{2}\nabla^{2}P(\mathbf{r},t)(\mathbf{r}'-\mathbf{r})^{2}.
\end{aligned}
\end{equation}
After substitution and keeping terms up to second order, the first-order term vanishes due to isotropy of the jump length distribution, which implies zero mean displacement and no drift. Eq.~\ref{differentiate_gme} then reduces to
\begin{equation}
\begin{aligned}
\frac{\partial P(\mathbf{r},t)}{\partial t}=D\nabla^{2}P(\mathbf{r},t), \qquad
D=\frac{\lambda\langle \ell^{2}\rangle}{2d}=\frac{\langle \ell^{2}\rangle}{2d\langle \tau\rangle},
\label{eq:diffusion_limit}
\end{aligned}
\end{equation}
where $d$ is the spatial dimension. Thus, for a decoupled CTRW under standard assumptions of homogeneity and finite moments, coarse-graining of the dynamics reduces to the ordinary diffusion equation, Eq.~\ref{eq:diffusion_limit}.

\subsection{Dipolar hopping model for spin polarization}

We now specialize the CTRW to nuclear spin transport. For a fixed configuration of $^{13}$C nuclei, let $W_{ij}$ with $i\ne j$ denote the rate for polarization to hop from spin $i$ to spin $j$. If $p_i$ denotes the polarization of spin $i$, the corresponding site-resolved polarization obeys the classical master equation
\begin{equation}
\begin{aligned}
\frac{dp_i(t)}{dt}=\sum_{j\ne i} W_{ji}p_j(t)-\lambda_i p_i(t), \qquad
\lambda_i=\sum_{j\ne i}W_{ij}.
\label{eq:site_master}
\end{aligned}
\end{equation}
If $\mathbf{p}=(p_1,p_2,\dots)$ is the site-resolved polarization vector, Eq.~\ref{eq:site_master} may be written as $d\mathbf{p}/dt=\mathcal{L}\mathbf{p}$, where $\mathcal{L}$ is the Markov generator. We reserve $\mathbf{W}$ for the off-diagonal rate matrix with elements $W_{ij}$. The generator has off-diagonal elements $\mathcal{L}_{ij}=W_{ji}$ for $i\ne j$ and diagonal elements $\mathcal{L}_{ii}=-\lambda_i$. The off-diagonal elements of $\mathcal{L}$ describe arrivals from other sites, analogous to the gain term in Eq.~\ref{differentiate_gme}, while the diagonal elements enforce probability conservation and describe escape from site $i$. 

The hopping rates are obtained from the secular dipolar coupling between like nuclei. We take
\begin{equation}
\begin{aligned}
W_{ij}=\mathcal{A}\,d_{ij}^{2}, \qquad i\ne j,
\label{eq:dipolar_rate}
\end{aligned}
\end{equation}
where the proportionality constant $\mathcal{A}$ absorbs the spectral overlap associated with flip-flop transitions~\cite{abragamPrinciplesNuclearMagnetism1983, ernstSpinDiffusionSolids1998, vonwitteNuclearSpinDiffusion2025}. A derivation of this expression and an estimate of $\mathcal{A}$ are provided in Appendix A. This estimate gives a value of $\mathcal{A}$ consistent with the value previously obtained by fitting experimental relaxation data in Ref.~\cite{selcoEmergentDecoherenceDynamics2025a}. While the value of $\mathcal{A}$ sets the overall timescale of the dynamics, the qualitative features of the transport behavior discussed below, and the main conclusions of this work, are insensitive to variations in this prefactor. The dipolar coupling is
\begin{equation}
\begin{aligned}
d_{ij}=-\frac{\hbar\mu_{0}}{4\pi}\frac{\gamma_C^{2}}{r_{ij}^{3}}\frac{1}{2}\left(3\cos^{2}\theta_{ij}-1\right),
\label{eq:dipolar_coupling}
\end{aligned}
\end{equation}
where $\mu_0$ is the vacuum permeability, $\gamma_C$ is the $^{13}$C gyromagnetic ratio, $r_{ij}$ is the separation between spins $i$ and $j$, and $\theta_{ij}$ is the angle between the inter-spin vector and the external magnetic field, taken here to be along the $z$-axis. Apart from the angular factor, Eq.~\ref{eq:dipolar_rate} together with Eq.~\ref{eq:dipolar_coupling} gives $W_{ij}\propto r_{ij}^{-6}$, which is the microscopic origin of the broad hierarchy of escape rates in a dilute, randomly occupied lattice.

\subsection{Dilute disordered spin network}

In this work, we investigate transport on a dilute lattice, where the random removal of lattice sites produces a positionally disordered network of occupied sites. We consider natural-abundance diamond, where the $^{13}$C concentration is 1.1\%, resulting in a dilute spin-$1/2$ network. The random occupation of lattice sites by $^{13}$C nuclei generates strongly inhomogeneous local environments, with some nuclei residing in densely connected clusters while others remain effectively isolated, leading to a broad distribution of transport pathways and timescales. We note that, throughout this work, the external magnetic field is taken along the $[100]$ crystal axis, consistent with previous experimental studies~\cite{beatrezElectronInducedNanoscale2023d, selcoEmergentDecoherenceDynamics2025a, selcoBreakdownDisorderSuppressedFloquet2026, shahTunableMpembaEffect2026}. For this orientation, the secular dipolar coupling between nearest-neighbor $^{13}$C (separated by $1.54$ \r{A}) vanishes due to the angular factor ($3\cos^2\theta-1=0$). Periodic boundary conditions are imposed to eliminate boundary effects and approximate bulk transport. Unless stated otherwise, paramagnetic impurities (nitrogen-vacancy (NV) centers and substitutional-nitrogen (P1) centers) are excluded in this part of the calculation so that the intrinsic nuclear transport can be isolated; electron-induced relaxation is introduced separately in Sec.~V.

\begin{figure}[t]
    \centering
    \includegraphics[width=\columnwidth]{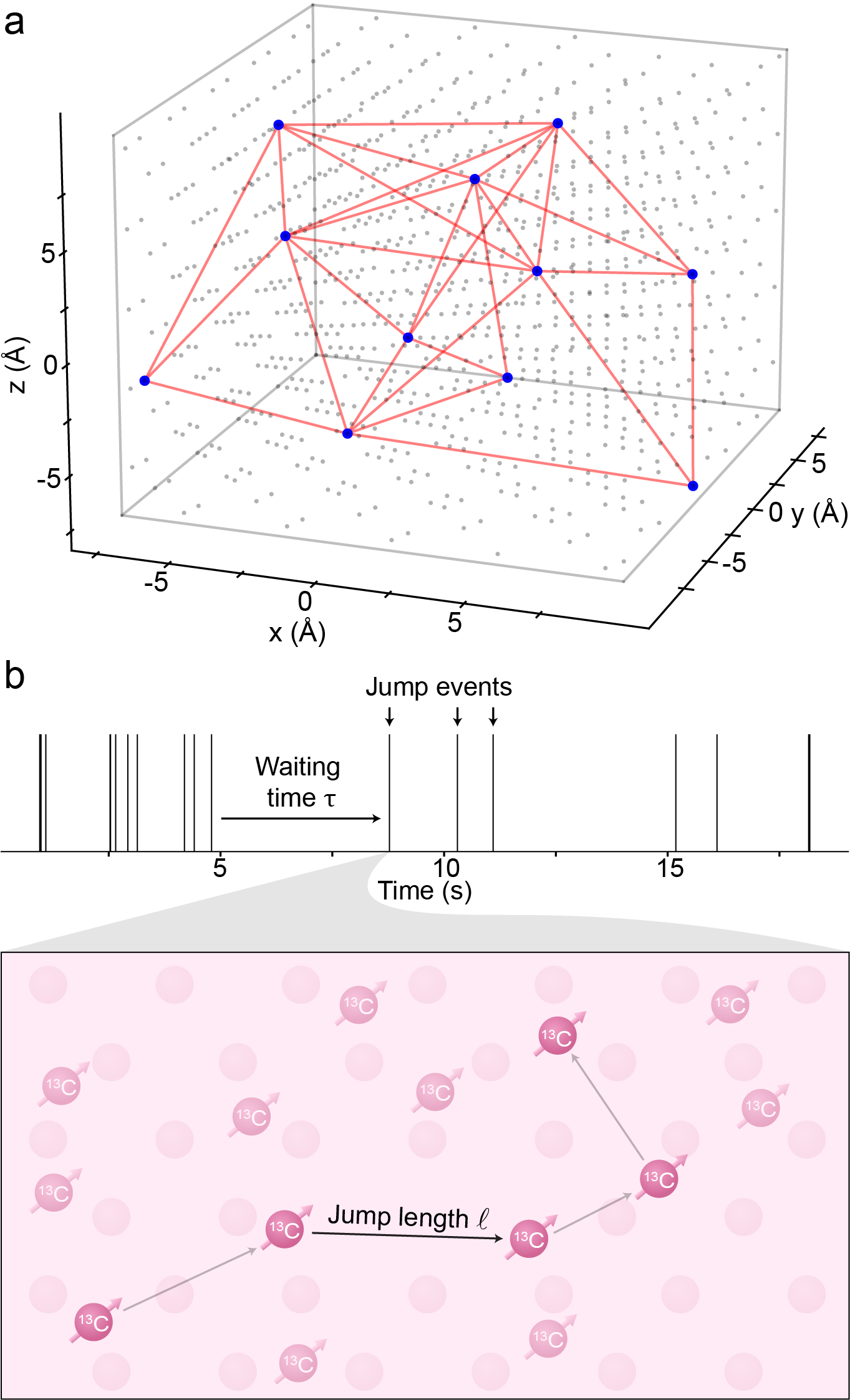}
    \caption{(a) Single random realization of the dilute $^{13}$C network in diamond. Gray points denote carbon lattice sites and blue points denote randomly occupied $^{13}$C sites at 1.1\% natural abundance. Red lines connect $^{13}$C pairs separated by less than $12~\mathrm{\AA}$, highlighting nearby local links; all off-diagonal dipolar rates are retained in the simulations. (b) Top: stochastic trajectory represented as a sequence of jump events in continuous time. The intervals between events are waiting times $\tau$. Bottom: schematic of one polarization hop between two $^{13}$C nuclei separated by a jump length $\ell$.}
    \label{Fig1}
\end{figure}

A representative realization of the $^{13}$C network is shown in Fig.~\ref{Fig1}a. $^{13}$C nuclei (blue points) randomly occupy diamond lattice sites (gray points) at a concentration of 1.1\%. Red lines mark pairs separated by less than $12~\mathrm{\AA}$ to indicate nearby local links; the simulations, however, use the full set of dipolar rates between all occupied sites. The resulting graph is formally connected by long-range couplings, but its weights are highly heterogeneous. This distinction between topological connectivity and kinetic accessibility is central to the anomalous transport discussed below.

\subsection{Stochastic trajectory simulation}

For each fixed disorder realization, Eq.~\ref{eq:site_master} defines a continuous-time Markov jump process. We sample this process using the Gillespie algorithm, which generates exact trajectories of the master equation without time discretization~\cite{gillespieGeneralMethodNumerically1976,gillespieExactStochasticSimulation1977}. A trajectory begins on a randomly selected $^{13}$C site. If the walker is currently on site $i$, the total escape rate is
\begin{equation}
\begin{aligned}
\lambda_i=\sum_{j\ne i}W_{ij},
\label{eq:gillespie_escape_rate}
\end{aligned}
\end{equation}
consistent with Eq.~\ref{eq:site_master}. The residence time on that site is drawn from
\begin{equation}
\begin{aligned}
\psi_i(\tau)\equiv\psi(\tau|\lambda_i)=\lambda_i e^{-\lambda_i\tau},
\label{eq:gillespie_waiting}
\end{aligned}
\end{equation}
where the exponential form follows from the assumption that the local hopping dynamics are Markovian. The destination site $j$ is then selected with probability
\begin{equation}
\begin{aligned}
\Pr(i\rightarrow j)=\frac{W_{ij}}{\lambda_i}.
\label{eq:gillespie_transition_prob}
\end{aligned}
\end{equation}
The physical time is advanced by $\tau$, and the procedure is repeated until the desired final time $T_{\rm max}$ is reached. The event sequence is summarized schematically in Fig.~\ref{Fig1}b. If the accumulated time first exceeds $T_{\rm max}$ during a waiting interval, that waiting interval is allowed to complete and the subsequent jump is executed before terminating the trajectory. This convention preserves the event-resolved waiting-time statistics.

For a fixed rate matrix, averages over many such trajectories are referred to as ensemble averages. We then average these quantities over independent random placements of the $^{13}$C nuclei, which we refer to as configurational averages. The number of spins, trajectories, and disorder realizations is varied with the observable being estimated. Typical calculations use $\sim10^{3}$--$10^{4}$ spins per disorder realization, $\sim10^{3}$ trajectories per realization, and $10^{2}$--$10^{3}$ independent disorder realizations. Specific parameters for each calculation are provided where relevant.

\section{Anomalous transport in the dilute $^{13}$C network}

\subsection{Waiting-time statistics and anomalous transport}

For a fixed disorder realization, the residence time on site $i$ is exponentially distributed with rate $\lambda_i$, as in Eq.~\ref{eq:gillespie_waiting}. Nontrivial behavior appears only after the walk samples the broad set of local escape rates generated by the dilute lattice. Because these escape rates are broadly distributed, the dynamics need not be effectively ergodic over experimentally relevant observation windows. The relevant waiting-time distribution is therefore not a single-rate exponential, nor simply a steady-state distribution associated with a fully mixed Markov chain; it is the empirical trajectory-level distribution obtained by recording every dwell time $\tau$ along many Gillespie trajectories and then averaging over trajectories and independent $^{13}$C configurations. This is the quantity most directly relevant for transport on experimental timescales, where the walker may not sample all regions of the network uniformly.

Fig.~\ref{Fig2}a shows the waiting-time distribution as a density per logarithmic interval, $\psi(\rm{log}\space\tau)$; for
natural logarithms, $\psi(\rm{log}\space\tau)=\tau\psi(\tau)$. This representation weights each decade of waiting times equally in log space and is therefore well suited for visualizing behavior across many orders of magnitude. The distribution begins at the fastest simulated events, on the order of $10$~ns, rises over several decades, reaches a broad maximum near $\tau\sim0.1$~s, and then crosses over to a slowly decaying tail. The fastest events are associated with strongly coupled local pairs and clusters, although in this regime coherent dipolar dynamics can become important. As discussed in Sec.~III~D, coarse graining over these clusters (Appendix~B) removes these ultrashort waiting time events while preserving the long-time transport behavior and thus does not affect the main conclusions of this work. The tail reflects rare, isolated spins that are weakly connected to the rest of the network.

The tail region encapsulates a regime where rare but dynamically significant events dominate. The ordinary probability density $\psi(\tau)$ in this tail regime is shown in Fig.~\ref{Fig2}b for $\tau>0.1$~s. As shown by the red line in Fig.~\ref{Fig2}b, it is well fit by
\begin{equation}
\begin{aligned}
\psi(\tau)=C_{\rm tail}\,\tau^{-(1+\alpha)}e^{-\tau/t_{\rm cutoff}},
\label{eq:waiting_tail}
\end{aligned}
\end{equation}
where $C_{\rm tail}$ is a fitted prefactor, with $\alpha=0.64$ and $t_{\rm cutoff}=19$~s.

In the absence of a cutoff, values $0<\alpha<1$ correspond to a heavy-tailed distribution which possesses a divergent mean waiting time~\cite{bouchaudAnomalousDiffusionDisordered1990, klafterFirstStepsRandom2011a}. Although the exponential cutoff renders the mean finite, the system behaves effectively as a CTRW with a heavy-tailed waiting-time distribution for times $t\ll t_{\rm cutoff}$. In this regime, the transport dynamics can exhibit signatures commonly associated with fractional kinetics, including aging and ergodicity breaking~\cite{metzlerRandomWalksGuide2000}. Thus, even though the microscopic transitions are purely Markovian, coarse-grained transport inherits anomalous characteristics from the broad distribution of local escape rates generated by positional disorder.

The presence of an exponential cutoff has a clear geometric origin. In the random placement of $^{13}$C on the diamond lattice, extremely weakly coupled spins arise from rare large separations, but the probability of arbitrarily large voids is strongly suppressed. Thus, there is a characteristic largest isolation scale which naturally produces a cutoff in the waiting-time distribution.

From the waiting-time statistics alone, one might anticipate that transport should ultimately cross over to normal diffusion for times $t\gg t_{\rm cutoff}$. As we show below, however, this expectation is incomplete: geometric trapping and connectivity constraints give rise to persistent sub-diffusive behavior that cannot be understood solely from the waiting-time distribution.

\subsection{Spatiotemporal coupling and concentration dependence}

The standard diffusion limit derived in Sec.~II assumes that spatial and temporal increments can be separated, so that the joint CTRW distribution factorizes as $\chi(\mathbf{r},\tau)=\phi(\mathbf{r})\psi(\tau)$. The dilute dipolar network violates this assumption at long dwell times. To quantify this effect, we compute the conditional mean jump length $\langle\ell\rangle_{\tau}\equiv\langle |\mathbf{r}|\mid \tau\rangle$, defined as the mean length of jumps that occur immediately after a waiting time $\tau$.

\begin{figure}[htbp!]
    \centering
    \includegraphics[width=\columnwidth]{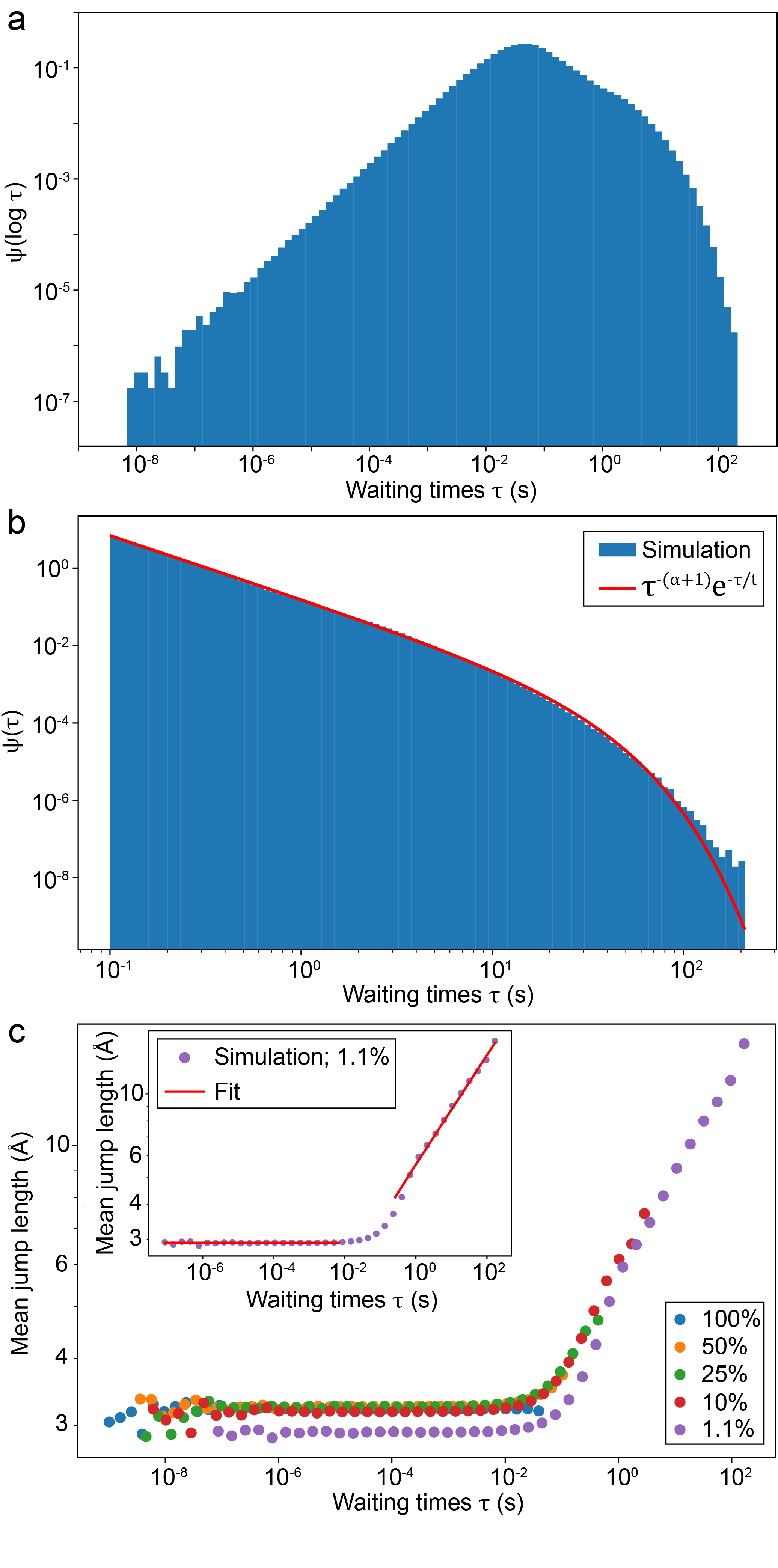}
    \caption{(a) Waiting-time distribution plotted as $\psi(\rm {log}\space\tau)$, the probability density per logarithmic interval. This representation weights each decade in time equally and highlights behavior across multiple orders of magnitude. (b) Probability density $\psi(\tau)$ on a log--log scale for $\tau>0.1$~s, emphasizing the tail beyond the peak of the distribution. The red line is a fit to $C_{\rm tail}\tau^{-(1+\alpha)}e^{-\tau/t_{\rm cutoff}}$, with $\alpha=0.64$ and $t_{\rm cutoff}=19$~s. (c) Mean jump length $\langle \ell \rangle_{\tau}$ conditioned on waiting time $\tau$. Inset: 1.1\% $^{13}$C data and a piecewise fit that is constant for $\tau\lesssim0.1$~s and scales as $\langle\ell\rangle_{\tau}\propto\tau^{0.20}$ for $\tau\gtrsim0.1$~s. Main panel: concentration dependence for 1.1\%, 10\%, 25\%, 50\%, and 100\% $^{13}$C. The coupled long-waiting-time regime becomes progressively less observable with increasing $^{13}$C concentration. Simulations used $\approx1100$ spins per realization, $10^{3}$ trajectories per disorder realization, and $10^{3}$ independent disorder realizations.}
    \label{Fig2}
\end{figure}

The inset of Fig.~\ref{Fig2}c shows $\langle\ell\rangle_{\tau}$ (purple points). For $\tau\lesssim0.1$~s, $\langle\ell\rangle_{\tau}$ is nearly constant, indicating that the distance traveled in a hop is effectively independent of the preceding waiting time and hence the walk is decoupled in this regime. For $\tau\gtrsim0.1$~s, however, the conditional mean grows as
\begin{equation}
\begin{aligned}
\langle \ell \rangle_{\tau}\propto\tau^{\beta},
\label{eq:jump_waiting_coupling}
\end{aligned}
\end{equation}
with $\beta=0.20$. The piecewise fit (red line) in the inset of Fig.~\ref{Fig2}c highlights the crossover from a locally decoupled hopping regime, where $\langle\ell\rangle_{\tau}$ is approximately constant, to a coupled CTRW regime in which longer residence times preferentially precede longer jumps.

The main panel of Fig.~\ref{Fig2}c further displays this behavior as a function of $^{13}$C concentration. Increasing the $^{13}$C enrichment increases the number of nearby hopping partners, suppresses rare isolated sites, and reduces the characteristic isolation scale discussed above. Consequently, $t_{\rm cutoff}$ decreases and the tail of the waiting-time distribution shrinks as the concentration increases, making dwell times beyond $\sim0.1$~s progressively less common. Therefore, at sufficiently high concentrations, long waiting times beyond $\sim0.1$ s are exceedingly rare, and the coupled regime becomes unobservable. In the dense limit, the walk therefore approaches the decoupled condition assumed by conventional spin-diffusion theory.

\subsection{Mean squared displacement and non-Fickian scaling}
We next return to the natural-abundance network and ask whether the anomalous waiting-time and jump statistics translate into anomalous spatial spreading. For an unbiased random walk with independent, finite-variance increments, the mean squared displacement (MSD) is expected to grow linearly with the number of steps, $\langle r^2(n)\rangle\propto n$, and ordinary Fickian diffusion further gives $\langle r^2(t)\rangle=6Dt$ in three dimensions. To separate long-time transport from the cutoff of the waiting-time tail, trajectories were propagated to times beyond $t_{\rm cutoff}$.

Fig.~\ref{Fig3}a shows the MSD as a function of step number $n$. Rather than growing linearly, the data follow
\begin{equation}
\begin{aligned}
\langle r^{2}(n) \rangle=An^{\gamma},
\label{eq:msd_steps}
\end{aligned}
\end{equation}
with $A=52.47~\mathrm{\AA}^{2}$ and $\gamma=0.56$. The gray dashed line in Fig.~\ref{Fig3}a shows the linear scaling expected for a normal walk in step space. The much smaller exponent indicates that many jump events do not lead to commensurate spatial exploration. In other words, the walker often revisits a local region many times before reaching new parts of the network, a hallmark of transport on heterogeneous or fractal-like connectivity graphs~\cite{klafterFirstStepsRandom2011a, hughesRandomWalksRandom1996}. This behavior anticipates the kinetic-percolation picture developed in Sec.~IV: although the long-range dipolar graph is formally connected, sparse bottlenecks between locally dense regions can strongly suppress spatial exploration per step.

The MSD in physical time is shown in Fig.~\ref{Fig3}b. It is also sublinear, but less strongly so, and is fit by
\begin{equation}
\begin{aligned}
\langle r^{2}(t) \rangle=6D_{\delta} t^{\delta},
\label{eq:msd_time}
\end{aligned}
\end{equation}
with $D_{\delta}=5.22~\mathrm{\AA}^{2}\mathrm{s}^{-\delta}$ and $\delta=0.87$, consistent with earlier analysis of this diamond system~\cite{selcoEmergentDecoherenceDynamics2025a}. The gray dashed line in Fig.~\ref{Fig3}b indicates the corresponding linear-in-time diffusive scaling. The inequality $\delta<1$ shows that the transport remains non-Fickian even at times longer than $t_{\rm cutoff}$. 

At the same time, $\delta>\gamma$ reflects the spatiotemporal coupling identified in Fig.~\ref{Fig2}c: long waiting times are statistically correlated with long spatial jumps. When the walker remains on a given site for an extended period, it is typically because the escape occurs through a weak but long-range coupling. Thus, although the MSD grows slowly as a function of step number due to repeated local hopping, the temporal evolution is partially ``boosted'' by the fact that rare, long waiting events are associated with disproportionately large displacements. In this sense, long waiting times are effectively rewarded with long jumps, which enhances the growth of the MSD in the time domain relative to step space. The difference between $\gamma$ and $\delta$ therefore directly reflects the non-separability of space and time in this transport process. The sub-diffusion observed here cannot be attributed solely to a heavy-tailed waiting time distribution nor solely to geometric trapping, but instead emerges from their nontrivial interplay within the long-range dipolar network. 

\subsection{Microscopic origin: geometric trapping}

\begin{figure}[htbp!]
    \centering
    \includegraphics[width=\columnwidth]{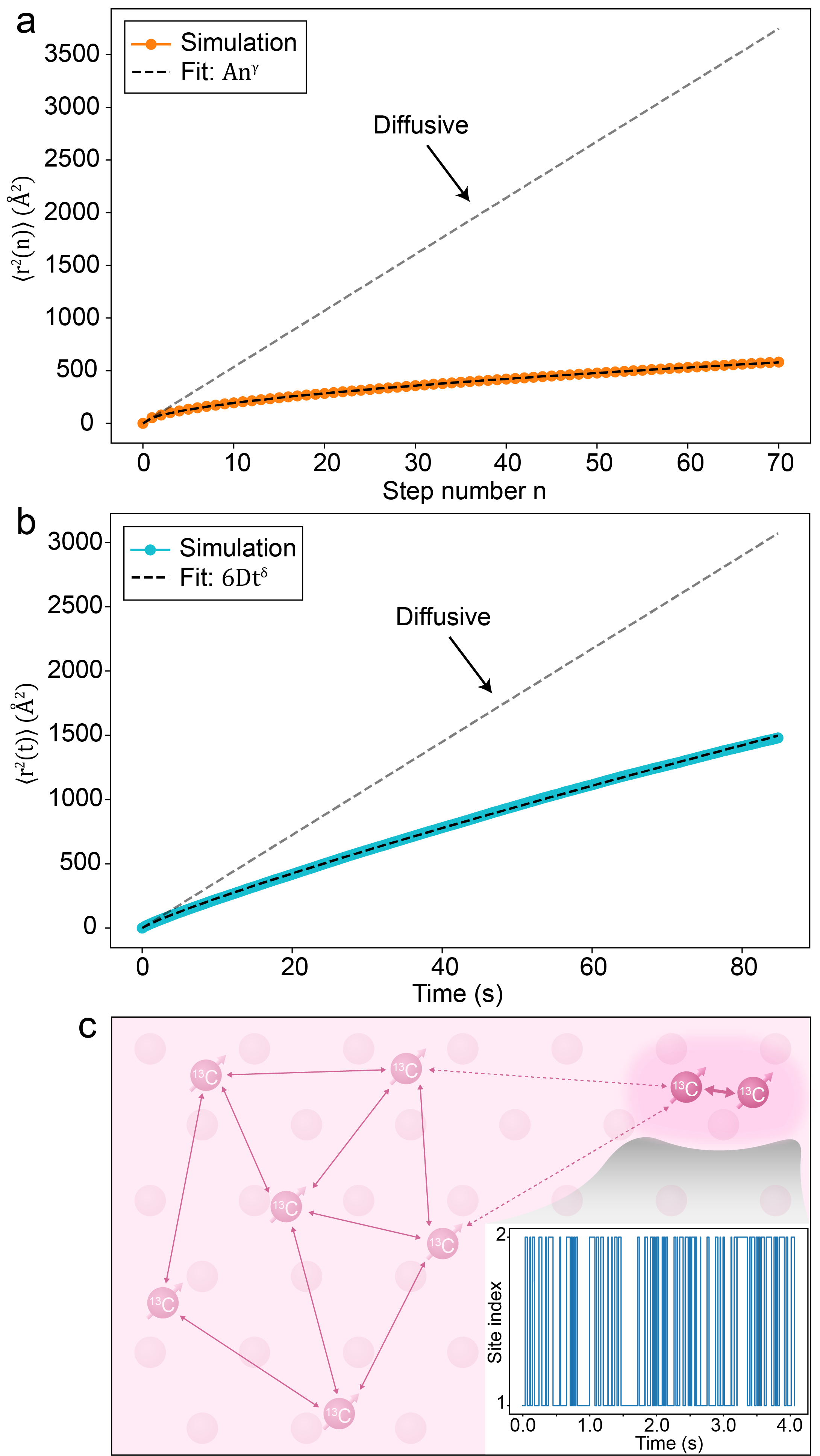}
    \caption{(a) Mean-squared displacement (MSD) as a function of step number $n$. The black dashed line is a fit to $An^{\gamma}$ with $\gamma=0.56$, while the gray dashed line indicates the linear scaling expected for a normal walk, $\langle r^{2}(n)\rangle\propto n$. (b) MSD as a function of physical time. The black dashed line is a fit to $6D_{\delta}t^{\delta}$ with $D_{\delta}=5.22~\mathrm{\AA}^{2}\mathrm{s}^{-\delta}$ and $\delta=0.87$; the gray dashed line shows the linear-in-time scaling expected for Fickian diffusion. Simulations used $\approx3200$ spins per realization, $10^{3}$ trajectories per disorder realization, and $10^{2}$ independent disorder realizations. (c) Schematic of geometric trapping in the dilute $^{13}$C network. A strongly coupled pair, shown as the highlighted dimer, supports repeated local exchange before escape to the surrounding network. The inset shows a representative simulated trajectory in which polarization repeatedly hops between two sites.}
    \label{Fig3}
\end{figure}

The sub-diffusion in step-space points to a microscopic mechanism: geometric trapping within small, strongly connected regions of the disordered network. Fig.~\ref{Fig3}c illustrates the simplest such structure, a strongly coupled pair embedded in a more weakly connected environment. Because the internal rate of the pair can greatly exceed the rates connecting either spin to the rest of the network, polarization can remain confined within the pair for many successive hops before escaping. The inset of Fig.~\ref{Fig3}c shows this behavior in a simulated trajectory, where the site index alternates repeatedly between two spins while the real-space displacement remains small. 
Consequently, the walker can accumulate many steps while remaining spatially localized, suppressing the growth of the MSD in step number.

This dimer picture should be viewed as the minimal example of a broader trapping phenomenon. More generally, local clusters with large internal $W_{ij}$ and weak external links act as kinetic reservoirs. They generate many hops, but only a small fraction of those hops advance the polarization front. Similar effects were recognized in early theories of rare-spin diffusion: Vugmeister analyzed positional disorder in the generator equation $\dot{\mathbf{p}}=\mathcal{L}\mathbf{p}$ and derived disorder-averaged effective diffusion constants~\cite{vugmeisterSpatialSpectralSpin1976c}, while Goldman and Jacquinot emphasized that repeated back-and-forth motion in strongly coupled pairs contributes little to net transport and introduced an efficiency coefficient to account for this loss~\cite{goldmanNuclearSpinDiffusion1982b}. The present trajectory-resolved CTRW formulation extends this picture by showing that the same microscopic motif produces a full set of non-Fickian statistics, including the sublinear MSD in Fig.~\ref{Fig3}a, the coupled jump-waiting statistics in Fig.~\ref{Fig2}c, and the anomalous time exponent in Fig.~\ref{Fig3}b.

\begin{figure}[t]
    \centering
    \includegraphics[width=\columnwidth]{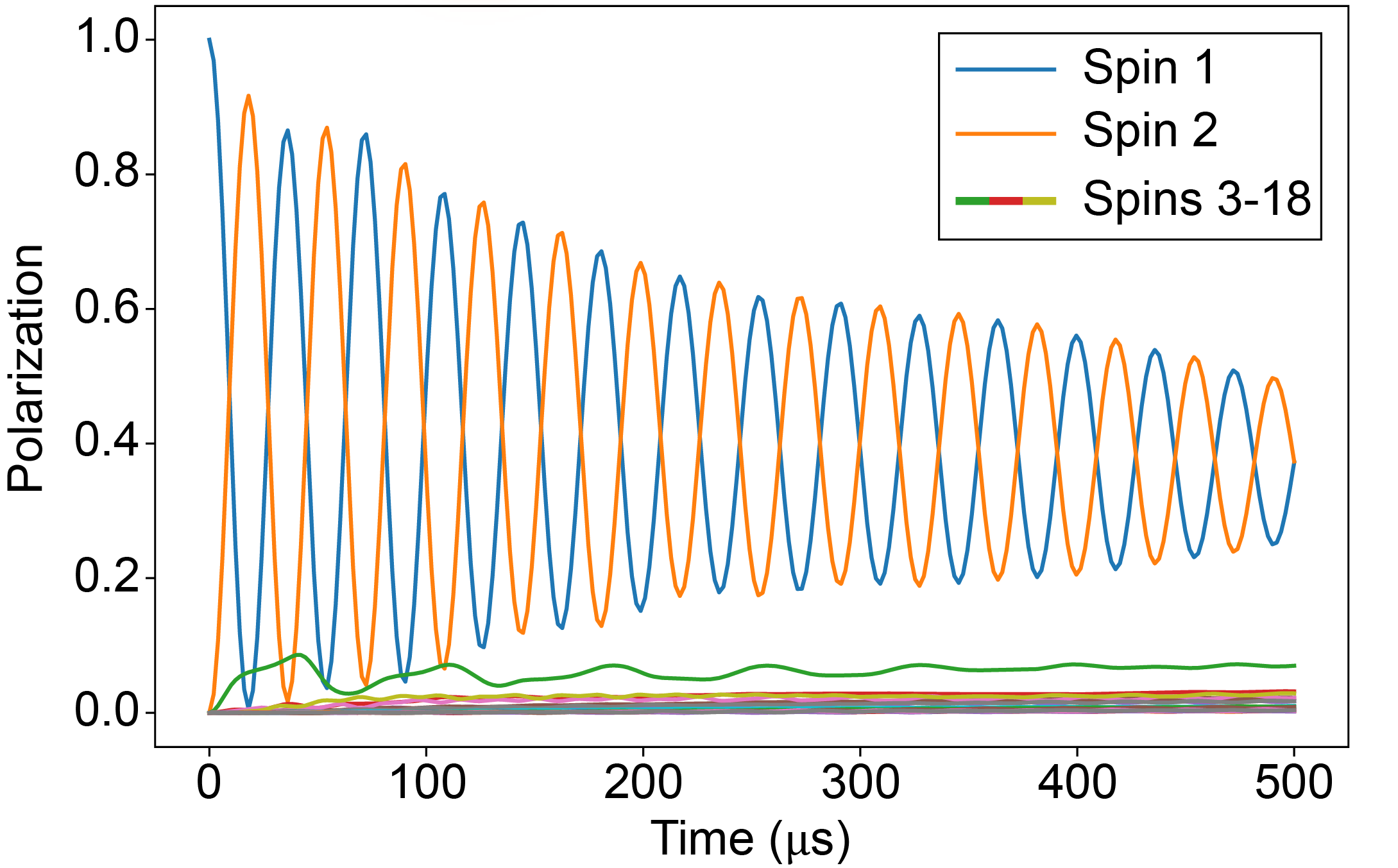}
    \caption{Fully quantum simulation of an 18-spin system with dipolar couplings corresponding to a random 1.1\% $^{13}$C configuration in diamond, initialized with polarization localized on spin 1. Site-resolved quantum dynamics show persistent oscillations between spins 1 and 2, while spins 3--18 acquire polarization only slowly, indicating trapping in the strongly coupled pair.}
    \label{Fig4}
\end{figure}

We note that within strongly coupled clusters, the incoherent hopping approximation used in the CTRW description may break down. In this regime, polarization exchange is more accurately described by coherent dipolar dynamics, rather than by an incoherent Poisson process. To test whether the trapping mechanism identified above is an artifact of the incoherent hopping approximation, we also simulated the coherent dynamics of small disordered clusters. Specifically, we performed fully quantum simulations of an 18-spin system with coupling topology drawn from random 1.1\% $^{13}$C configurations in diamond. For each configuration, the initial polarization was localized on spin 1, and the resulting unitary evolution of the site-resolved polarization was tracked. As shown in Fig.~\ref{Fig4}, the dynamics exhibit pronounced oscillations between spin 1 and a nearby spin 2, while polarization spreads only slowly to the remaining sites; similar behavior is observed across other disorder realizations. Thus, the same geometric structures that trap trajectories in the CTRW picture also confine polarization in the fully quantum dynamics. 


Importantly, we note here that the precise microscopic description of these short-time intra-cluster dynamics does not determine the anomalous transport behavior discussed in this work. Rather, the relevant transport behavior emerges on longer timescales, after rapid local equilibration within strongly coupled clusters. As shown in Appendix~B, we construct a coarse-grained description of the CTRW process in which each strongly coupled cluster is treated as a single effective state. The resulting escape dynamics between clusters are well described by an effective Poisson process, and this coarse-grained model reproduces the long-time tail of the waiting-time distribution (Fig.~\ref{Fig2}b) and the MSD scaling with time (Fig.~\ref{Fig3}b). Consequently, the anomalous transport discussed in this work is governed by the broad distribution of inter-cluster escape times rather than by the precise microscopic dynamics within individual clusters.


The physical picture that emerges is therefore one of local equilibration followed by rare escape. Strong links rapidly redistribute polarization within small clusters, while weak links control the rate at which polarization reaches new spatial regions. Sec.~IV turns this picture into a kinetic percolation analysis of the weighted rate matrix $\mathbf{W}$.

\section{Kinetic percolation approach}

\begin{figure*}[t]
    \centering
    \includegraphics[width=\textwidth]{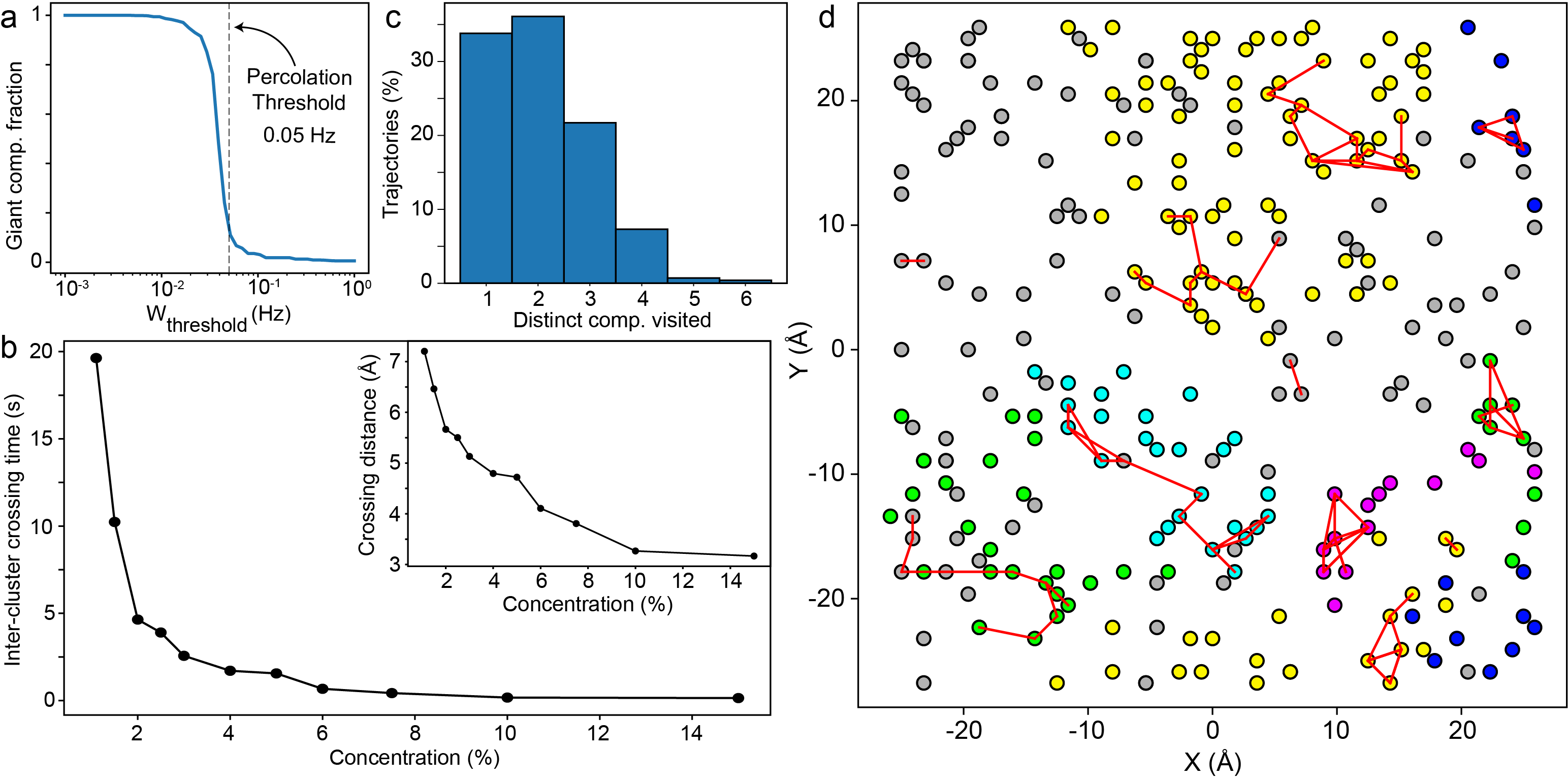}
    \caption{Kinetic percolation analysis of the weighted $^{13}$C hopping network. (a) Fraction of sites in the largest connected component (``giant component'') of the thresholded graph $B_{ij}(W_{\rm thresh})$ for a representative 1.1\% $^{13}$C configuration. The rapid drop near $W_{\rm thresh}\approx0.05$~Hz signals fragmentation of the network into smaller clusters and defines the operational kinetic-percolation scale $W_{\rm perc}$. (b) Inter-cluster crossing time $\tau_{\rm cross}=W_{\rm perc}^{-1}$ as a function of $^{13}$C concentration. Inset: corresponding crossing distance inferred from the dipolar rate neglecting angular dependence. Each point is obtained by averaging $W_{\rm perc}$ over $10^{2}$ independent disorder realizations. (c) Distribution of the number of distinct $B^{({\rm perc})}$ clusters visited by full-rate-matrix trajectories up to $\tau_{\rm cross}=20$~s at 1.1\% $^{13}$C. (d) Two-dimensional projection of a representative 1.1\% $^{13}$C configuration. The five largest $B^{({\rm perc})}$ clusters are colored, and red lines show example trajectories generated with the full weighted rate matrix $\mathbf{W}$, illustrating confinement within strongly coupled clusters even when all weak links are retained.}
    \label{Fig5}
\end{figure*}

The results of Sec.~III show that the anomalous walk is not produced by waiting-time statistics alone. The MSD is already sublinear when plotted against the number of jump events, Fig.~\ref{Fig3}a, which means that many microscopic hops do not produce commensurate exploration of new spatial regions. This behavior is reminiscent of random walks near a percolation threshold~\cite{klafterFirstStepsRandom2011a}. In its standard formulation, percolation theory assumes nearest-neighbor connectivity: a walker may hop only between adjacent occupied sites. As the occupation probability decreases, the system approaches a so-called percolation threshold at which the largest connected component collapses, producing a geometric phase transition beyond which global transport is no longer possible~\cite{havlinDiffusionDisorderedMedia1987, hughesRandomWalksRandom1996, staufferIntroductionPercolationTheory2018a}. The present system differs in an essential way: because the radial part of the dipolar rate scales as $r_{ij}^{-6}$, the weighted graph is long-ranged, rendering the network formally fully connected at any nonzero concentration. A walker can, in principle, reach any site given sufficient time, so no strict percolation transition exists in the conventional sense. The relevant question is therefore not whether an infinite cluster exists, but on what time scale the weak links between locally connected clusters become dynamically active.

We address this question by introducing a kinetic, rather than purely geometric, percolation construction. The construction is used as a diagnostic of the weighted rate matrix $\mathbf{W}$: it identifies groups of spins that are mutually connected by rates faster than a chosen timescale and separates them from links that are present in principle but too slow to control transport over that time window. 

\subsection{Kinetic percolation and network fragmentation}

For a fixed disorder realization, we choose a rate threshold $W_{\rm thresh}$ and define the binary connectivity matrix
\begin{equation}
\begin{aligned}
B_{ij}(W_{\rm thresh})=\begin{cases}
    1, & W_{ij}>W_{\rm thresh}, \\
    0, & W_{ij}\leq W_{\rm thresh},
\end{cases}
\label{eq:threshold_graph}
\end{aligned}
\end{equation}
for $i\ne j$, with $B_{ii}=0$. Edges retained in $B_{ij}$ correspond to polarization exchange events with characteristic times shorter than $W_{\rm thresh}^{-1}$, while weaker edges are treated as kinetically inaccessible on that time scale. Varying $W_{\rm thresh}$ therefore probes the connectivity of the network on different timescales: large $W_{\rm thresh}$ retains only the fastest local connections, whereas smaller values progressively incorporate slower links that become relevant at longer times.

Fig.~\ref{Fig5}a shows the size of the largest connected component of $B_{ij}(W_{\rm thresh})$, often referred to as the giant component. The size is normalized to unity for a fully connected network, and is plotted as a function of $W_{\rm thresh}$ for a representative $^{13}$C configuration at 1.1\% concentration. For small thresholds, nearly all $^{13}$C sites belong to a single connected component. As the threshold is increased, the largest connected component collapses sharply near
\begin{equation}
W_{\rm perc}\simeq 0.05~{\rm Hz},
\label{eq:wperc_value}
\end{equation}
indicating fragmentation of the network into smaller clusters. Repeating this analysis over many disorder realizations yields a consistent threshold near this value. We therefore identify $W_{\rm thresh}=0.05$ Hz as a characteristic ``kinetic percolation threshold'' and define the corresponding binary matrix as
\begin{equation}
\begin{aligned}
B^{({\rm perc})}_{ij}=\begin{cases}
    1, & W_{ij}>W_{\rm perc}, \\
    0, & W_{ij}\leq W_{\rm perc}.
\end{cases}
\label{eq:percolation_binary}
\end{aligned}
\end{equation}

\subsection{Kinetic bottlenecks and inter-cluster transport}

The rate scale in Eq.~\ref{eq:wperc_value} immediately gives a characteristic inter-cluster crossing time,
\begin{equation}
\tau_{\rm cross}=W_{\rm perc}^{-1}\approx 20~{\rm s}
\label{eq:crossing_time}
\end{equation}
for natural-abundance diamond. This time is essentially the same as the cutoff $t_{\rm cutoff}=19$~s extracted independently from the tail of the waiting-time distribution in Eq.~\ref{eq:waiting_tail}. The agreement suggests that the cutoff is set by the slowest geometrically relevant bridges between fast clusters: before this time, the walker predominantly samples local clusters; on comparable and longer times, rare inter-cluster escape events begin to contribute to long-range transport. This separation of timescales may be particularly important for polarizing large materials, where protocols such as dynamic nuclear polarization (DNP) must overcome such transport bottlenecks to enable efficient buildup of long-range polarization~\cite{priscoScalingAnalysesHyperpolarization2021a}.

Fig.~\ref{Fig5}b shows how this inter-cluster crossing time changes with $^{13}$C concentration. Increasing the concentration increases the number of nearby hopping partners and shifts the fragmentation of the network to larger rates, thereby reducing $\tau_{\rm cross}$. The inset of Fig.~\ref{Fig5}b expresses the same trend as an inter-cluster crossing distance, which is calculated by first converting $\tau_{\rm cross}$ into a hopping rate, $W_{\rm perc}=1/\tau_{\rm cross}$, then converting the hopping rate into a dipolar coupling using Eq.~\ref{eq:dipolar_rate}, and finally converting the coupling into a distance using Eq.~\ref{eq:dipolar_coupling}, neglecting the angular dependence of the dipolar interaction. At 1.1\% $^{13}$C this distance is about $7~\mathrm{\AA}$, whereas at higher enrichment it approaches $3~\mathrm{\AA}$. The latter value is consistent with the short-waiting-time jump length in the flat regime of Fig.~\ref{Fig2}c, where $\langle\ell\rangle_{\tau}$ is nearly independent of $\tau$. Thus, increasing $^{13}$C concentration collapses the distinction between local hops and inter-cluster hops, explaining why the enriched networks approach the decoupled, Fickian diffusion limit discussed in Secs.~II~A and III~B.

\subsection{Cluster-resolved transport and trajectory statistics}
The kinetic percolation construction is useful only if its clusters are reflected in the actual dynamics generated by the full rate matrix $\mathbf{W}$. To test this, we return to the 1.1\% $^{13}$C case and assign each site to a connected component of $B^{({\rm perc})}$, but random walk trajectories are propagated with all rates $W_{ij}$ retained. We then count the number of distinct clusters visited by each trajectory before $\tau_{\rm cross}=20$~s. The resulting distribution is shown in Fig.~\ref{Fig5}c. Most trajectories visit only one or two clusters, and approximately 90\% visit three or fewer clusters during the crossing-time window. This limited cluster exploration is the trajectory-level counterpart of the sublinear step-space MSD in Fig.~\ref{Fig3}a.

Fig.~\ref{Fig5}d provides a spatial view of the same effect, shown as a two-dimensional projection. The five largest components of $B^{({\rm perc})}$ are shown in different colors for a representative configuration, while the red paths are independent representative random-walk trajectories generated with the unthresholded rate matrix $\mathbf{W}$. The paths remain largely confined within one colored region and cross between colors only rarely. Because the red trajectories include all below-threshold couplings, this confinement is not imposed by the analysis; it is an emergent kinetic consequence of the broad distribution of $W_{ij}$. In some trajectories the walker remains within a single dimer or small cluster throughout the full 20~s interval, connecting the cluster-resolved picture in Fig.~\ref{Fig5}d to the dimer trapping mechanism of Fig.~\ref{Fig3}c and the coherent two-spin exchange shown in Fig.~\ref{Fig4}.

The kinetic percolation construction therefore provides a compact way to summarize the microscopic origin of anomalous transport. Strong links generate rapid equilibration within local clusters, while weak links set a concentration-dependent escape time, and the coexistence of these two timescales prevents the dilute network from being represented by a single, time-independent diffusion coefficient.

\section{Hard-sphere trapping and connection to experiment}

Secs.~III and IV considered intrinsic transport in the nuclear network, with paramagnetic impurities excluded. Experiments on nitrogen-doped diamond necessarily include such impurities, primarily NV centers and P1 centers, which provide spatially localized relaxation channels for $^{13}$C polarization~\cite{selcoEmergentDecoherenceDynamics2025a}. We now ask whether the anomalous transport influences the relaxation dynamics. To address this, we introduce a mapping that is closely related to classic static-trap reaction kinetics, in which a particle diffuses through a medium containing immobile absorbing hard-spheres~\cite{smoluchowskiDreiVortrageUber1916a,bixonDiffusionMediumStatic1981a,kirkpatrickTimeDependentTransport1982a,grassbergerLongTimeProperties1982b}. 

In the spin problem, the random walker is nuclear polarization and electron-induced relaxation plays a role analogous to trapping. Because the electron–induced relaxation scales as $1/r^6$~\cite{selcoEmergentDecoherenceDynamics2025a}, relaxation is quasi-localized in space. Spins sufficiently close to an electron experience rapid decay, while distant spins relax much more slowly. Although the analogy is not exact, we approximate this soft $r^{-6}$ sink landscape by a hard-sphere model: each electron is surrounded by a spherical trap of radius $R_{\rm trap}$, and a trajectory is terminated when it first reaches a $^{13}$C site inside any trap. Fig.~\ref{Fig6}a illustrates this construction. 

Electrons are included at a concentration of 30 ppm. To construct the trapping model, we randomly distribute both $^{13}$C spins and electrons at their respective concentrations. Random walks are then performed on the same dipolar network as before, with the modification that whenever a walker enters a spherical region of radius $R_{\rm trap}$ surrounding an electron, it is immediately annihilated. The resulting survival probability $S(t)$ of the walker serves as a proxy for the transport-assisted polarization decay.

\subsection{Determination of trap radius}
A key parameter in this construction is the choice of $R_{\rm trap}$. A useful reference scale is the spin-diffusion-barrier, or frozen-core, radius, which is approximately $16~\mathrm{\AA}$ for this diamond system~\cite{khutsishviliSpinDiffusion1966a,selcoEmergentDecoherenceDynamics2025a}. This provides a useful benchmark for assessing whether the resulting value of $R_{\rm trap}$ is physically reasonable. Rather than selecting $R_{\rm trap}$ arbitrarily, we determine it self-consistently from simulations that retain the explicit distance-dependent electron-induced relaxation.

In these calibration simulations, we use the semiclassical electron-induced relaxation model developed in Ref.~\cite{selcoEmergentDecoherenceDynamics2025a}. In that model, each nuclear site is assigned an effective on-site decay rate proportional to the sum of the squared dipolar couplings to the surrounding electron spins, along with a prefactor described in Ref.~\cite{selcoEmergentDecoherenceDynamics2025a}. For the present simulations, during each waiting event the accumulated decay is then given by the product of this effective on-site decay rate and the residence time. Each trajectory is propagated until 99\% of its polarization weight has decayed, and we record the minimum distance between the walker and any electron prior to annihilation. After ensemble and configurational averaging, the median minimum distance is approximately $20~\mathrm{\AA}$. This value is slightly larger than the frozen-core radius, providing a consistency check that the effective hard-sphere trapping radius is physically reasonable. We therefore set
\begin{equation}
R_{\rm trap}=20~\mathrm{\AA}.
\label{eq:rtrap_value}
\end{equation}
For 1.1\% $^{13}$C and 30~ppm electrons, this choice places roughly 13\% of the $^{13}$C sites inside trap regions.

\subsection{Early-time survival and heavy-tailed dynamics}

\begin{figure}[t]
    \centering
    \includegraphics[width=\columnwidth]{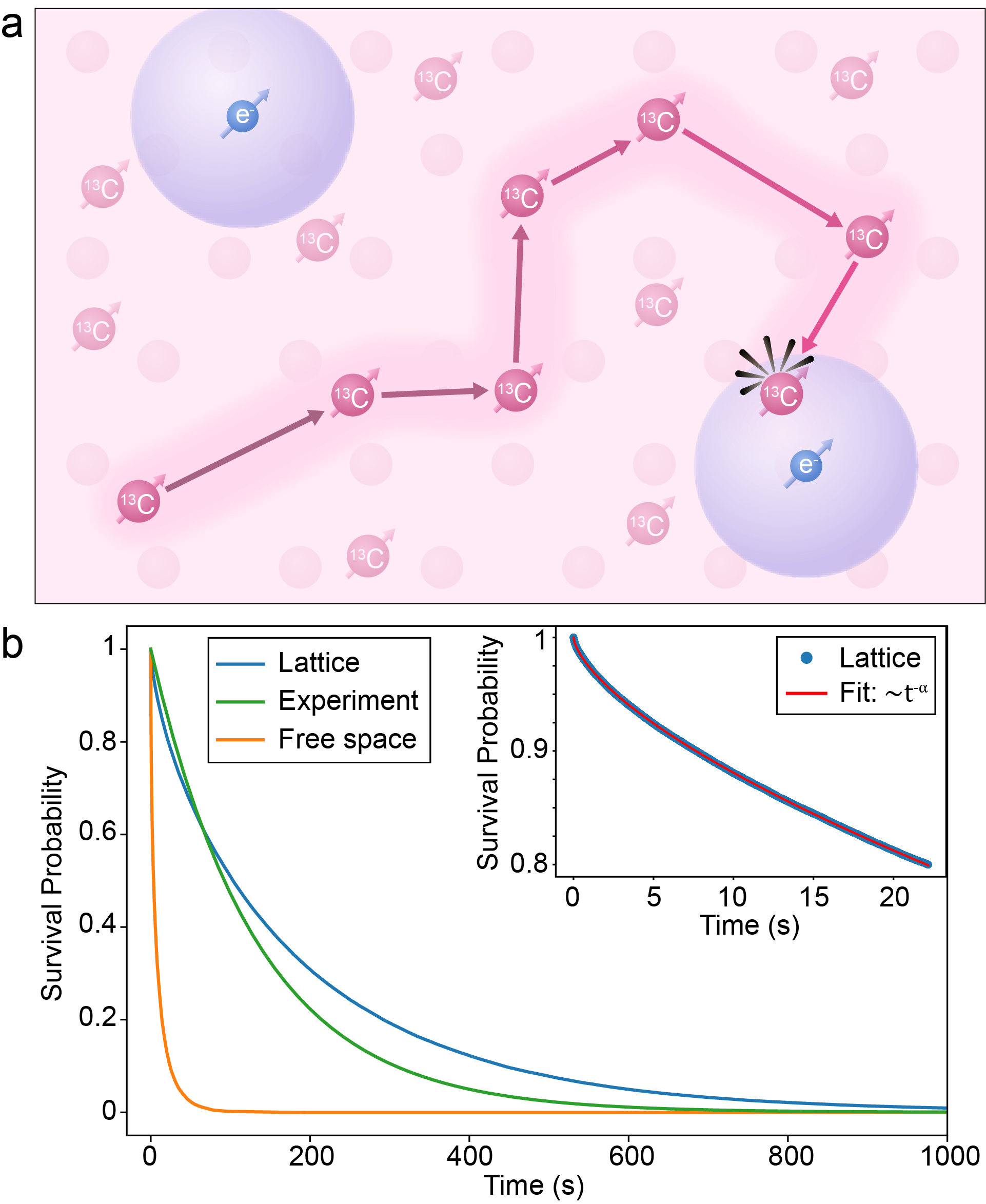}
    \caption{Hard-sphere trapping as a proxy for transport-mediated relaxation. (a) Schematic of the lattice trapping model. Blue spheres denote absorbing regions of radius $R_{\rm trap}$ around electron spins; when polarization hops to a $^{13}$C site inside a sphere, the trajectory is annihilated. This hard-sphere construction approximates the spatially localized $r^{-6}$ electron-induced relaxation. (b) Survival probability for 1.1\% $^{13}$C, 30~ppm electrons, and $R_{\rm trap}=20~\mathrm{\AA}$. Inset: early-time survival over the first $\sim20$~s, showing the theoretically expected scaling $S(t)\sim t^{-\alpha}$ with $\alpha=0.65$, consistent with the waiting-time exponent in Eq.~\ref{eq:waiting_tail}. Main panel: comparison of the lattice trapping simulation (blue), the experimentally measured transport-assisted relaxation contribution (green), and a continuous-space diffusion equation trapping simulation (orange). The lattice model qualitatively captures the experimental timescale, whereas the continuum diffusion equation model does not. Simulations used $\approx3200$ spins per realization, $10^{3}$ trajectories per disorder realization, and $10^{2}$ independent disorder realizations.}
    \label{Fig6}
\end{figure}

The early-time survival probability provides a direct test of whether the relaxation dynamics inherit the CTRW statistics of Sec.~III. For a low density of static traps and a heavy-tailed waiting-time distribution $\psi(\tau)\sim \tau^{-(1+\alpha)}$, the early-time survival probability is predicted to scale as
\begin{equation}
\begin{aligned}
S(t)\sim t^{-\alpha}
\label{eq:trap_survival_power}
\end{aligned}
\end{equation}
~\cite{klafterFirstStepsRandom2011a}. As shown in the inset of Fig.~\ref{Fig6}b, the simulated survival probability exhibits precisely this scaling during the first $\sim20$ s, with exponent $\alpha=0.65$, in close agreement with the value $\alpha=0.64$ obtained independently from the waiting-time distribution in Eq.~\ref{eq:waiting_tail}. This agreement connects the short-time relaxation dynamics directly to the heavy-tailed waiting time statistics that underlie the anomalous transport. While no closed-form analytical expression exists for the survival probability in three dimensions, asymptotic bounds can be derived~\cite{grassbergerLongTimeProperties1982b, klafterFirstStepsRandom2011a}.

\subsection{Comparison to experiment}

The main panel of Fig.~\ref{Fig6}b compares the lattice survival probability (blue) with the experimentally measured transport-mediated relaxation contribution (green) reported in Ref.~\cite{selcoEmergentDecoherenceDynamics2025a}. The agreement is qualitative, but it is significant: the hard-sphere model has replaced the true $r^{-6}$ electron relaxation by an instantaneous absorbing boundary, has neglected angular factors in the electron--nuclear coupling, and uses a single effective radius set by Eq.~\ref{eq:rtrap_value}. Despite these simplifications, the decay occurs on the correct experimental timescale, indicating that the experiment is sensitive to the same network-controlled transport process quantified by the CTRW statistics and the kinetic percolation analysis.

The comparison should not be interpreted as a complete microscopic relaxation theory. Quantitative agreement can be achieved by incorporating distance-dependent relaxation (“soft-spheres”) that retains the full $1/r^6$ distance dependence and angular structure of the electron-induced relaxation, as demonstrated in Ref.~\cite{selcoEmergentDecoherenceDynamics2025a}. The value of the hard-sphere construction is instead to capture the essential physics: it isolates the role of transport towards localized sinks in the relaxation process and shows that retaining the microscopic, disordered $^{13}$C hopping network is sufficient to qualitatively reproduce the observed relaxation timescale.

\subsection{Failure of the diffusion equation description}

To test whether the microscopic network geometry is necessary to reproduce the experimentally observed relaxation dynamics, we repeated the hard-sphere calculation in continuous three-dimensional space. In this comparison, the lattice walk is replaced by Fickian Brownian motion among the same density of absorbing spheres. The diffusion coefficient is chosen from the first two moments of the lattice CTRW, using the three-dimensional form of Eq.~\ref{eq:diffusion_limit},
\begin{equation}
\begin{aligned}
D=\frac{\langle \ell^{2} \rangle}{6\langle \tau \rangle},
\label{eq:effective_diffusion}
\end{aligned}
\end{equation}
where $\langle\ell^{2}\rangle$ and $\langle\tau\rangle$ are computed from the lattice CTRW trajectories used in the preceding analysis. 

As shown in Fig.~\ref{Fig6}b, the survival probability obtained from the diffusion-equation simulation (orange) deviates dramatically from both the lattice simulation and the experimental data. The discrepancy is not merely quantitative but qualitative: the overall timescale of the decay is fundamentally inconsistent.

One might argue that the diffusion coefficient could be tuned phenomenologically to improve agreement with experiment. Indeed, by selecting an appropriate value of $D$, it is possible to reproduce either the short- or the long-time behavior of the survival probability. However, no single constant diffusion coefficient yields agreement across the full time range. If $D$ is chosen to match the early-time dynamics, the late-time decay is substantially misrepresented, and vice versa.

This failure reflects a deeper issue. In anomalous transport, where the mean squared displacement scales as $\langle r^2(t) \rangle \sim t^{\delta}$ with $\delta \neq 1$, the effective diffusivity is not constant but evolves in time. Formally, one may define an instantaneous diffusivity $D_{\rm eff}(t)=\frac{1}{6}\frac{d}{dt} \langle r^2(t) \rangle$, which scales as $t^{\delta - 1}$. For $\delta < 1$, this quantity decreases with time, reflecting progressively slower spatial exploration. A classical diffusion equation assumes instead a time-independent transport coefficient, thereby enforcing linear growth of the MSD and a single transport timescale. It therefore cannot simultaneously capture both the early-time and late-time behavior of the lattice dynamics. In this sense, the time dependence of the effective transport coefficient is precisely the hallmark of fractional dynamics and aging: the system does not admit a stationary coarse-grained description, and its observables explicitly depend on the observation time.

The discrepancy is thus not merely quantitative but structural. The diffusion equation imposes a constant coarse-grained transport rate, whereas the dipolar network generates a temporally evolving effective diffusivity rooted in geometric trapping and heterogeneous connectivity. Only simulations that retain the full disordered hopping network preserve this time-dependent transport character and reproduce the experimentally relevant relaxation dynamics.

\section{Conclusions and outlook}
We have shown that polarization transport in dilute, positionally disordered spin networks is controlled by the geometry and connective topology of the underlying interaction graph. In the natural-abundance $^{13}$C network in diamond, the combination of positional disorder and $r^{-6}$ hopping rates produces broad waiting-time statistics, spatiotemporal correlations, and sub-diffusive growth of the mean squared displacement. The central microscopic mechanism is geometric trapping: strongly coupled local clusters, including dimers, repeatedly exchange polarization internally while weak links to the surrounding network control the much slower process of global exploration. 

The kinetic percolation analysis introduced here provides a compact way to translate this geometric picture into a dynamical timescale. Because dipolar interactions are long-ranged, the network is formally connected at any nonzero spin concentration; nevertheless, thresholding the weighted rate matrix reveals clusters that are effectively connected only on timescales longer than a characteristic inter-cluster crossing time. For natural-abundance diamond, this time is on the order of $\sim20$ s, consistent with the cutoff of the waiting-time distribution and to the timescale over which trajectories remain confined to a small number of clusters. This separation between fast intra-cluster exchange and slow inter-cluster motion is the key physical distinction between the dilute network studied here and the well-connected spin networks for which conventional diffusion theory is most reliable.

The hard-sphere trapping construction further connects this spin-transport problem to classic studies of reaction kinetics, where particles diffuse in media containing randomly distributed static traps~\cite{smoluchowskiDreiVortrageUber1916a, bixonDiffusionMediumStatic1981a, kirkpatrickTimeDependentTransport1982a, grassbergerLongTimeProperties1982b}. In this mapping, polarization plays the role of the random walker and paramagnetic impurities act as hard-sphere traps. The fact that the lattice trapping model captures the qualitative timescale of the experimental relaxation, while an effective continuum diffusion model does not, shows that relaxation measurements can be probes of microscopic transport heterogeneity. It also suggests that extracting a single diffusion coefficient from dilute-spin relaxation data can be misleading unless the observation window, trap geometry, and microscopic transport dynamics are explicitly accounted for.

Several extensions follow naturally. It will be useful to combine the present CTRW framework with coherent quantum dynamics beyond small clusters, in order to determine when interference, energy disorder, or many-body effects modify the classical hopping picture. Experimentally, the kinetic crossing time identified here suggests direct tests based on varying the observation window, isotope concentration, defect density, or Floquet-engineered hopping strength. Such measurements could resolve the crossover from local cluster equilibration to long-range polarization transport and could provide a route to engineering polarization flow in DNP and quantum sensing protocols. The findings presented here may also offer a useful perspective on transport through the spin-diffusion barrier surrounding paramagnetic defects, where the interplay between local confinement and transport through weakly connected pathways may influence the transfer of polarization to the bulk spin network~\cite{horvitzNuclearSpinDiffusion1971b, wolfeDirectObservationNuclear1973a, sternDirectObservationHyperpolarization2021b, chessariRoleElectronPolarization2023a, pangHypershiftedSpinSpectroscopy2024, vonwitteTwoelectronTwonucleusEffective2025}.

More broadly, these results emphasize that anomalous transport in long-range disordered networks can arise even when all microscopic transitions are memoryless. The apparent memory is generated by the quenched positional disorder of the network itself. Similar combinations of local trapping, weak inter-cluster links, and spatiotemporal coupling may occur in other dilute spin systems~\cite{appelbaumElectronicMeasurementControl2007, jansenSiliconSpintronics2012, zuEmergentHydrodynamicsStrongly2021b, wuSpinSqueezingEnsemble2025a}, in excitonic and charge transport through disordered molecular materials~\cite{coropceanuChargeTransportOrganic2007, akselrodVisualizationExcitonTransport2014, akselrodSubdiffusiveExcitonTransport2014}, and in synthetic quantum systems with sparse or long-range connectivity~\cite{schemppCorrelatedExcitonTransport2015a, trautmannTrappedionQuantumSimulation2018a}. A trajectory-based stochastic description therefore offers a useful bridge between microscopic disorder and emergent transport laws across a range of chemical and quantum materials.

\subsection*{Acknowledgements}
We acknowledge funding from the DOE (BES DE-SC0025524), NIH (1R01GM143626-01A1), ONR (N00014-24-1-2185), and AFOSR (FA9550-23-1-0106), as well as instrumental support from NSF (MRI 2320520). This material is based upon work supported by the Air Force Office of Scientific Research under award number FA9550-25-C-B010.

\subsection*{Appendix A: Estimation of the spectral overlap in the hopping rate}
Here we present a brief sketch of the derivation of the hopping-rate prefactor $\mathcal{A}$ in Eq.~\ref{eq:dipolar_rate} in the main text. For a more thorough derivation, see the Supplemental Information of Ref.~\cite{shahTunableMpembaEffect2026}. The truncated Hamiltonian describing flip-flop processes between spins $i$ and $j$ is
\begin{equation*}
H_{ij} = -\frac{1}{4}d_{ij}(I^{+}_{i}I^{-}_{j} + I^{-}_{i}I^{+}_{j}).
\end{equation*}
The effects of the surrounding spin bath are treated as stochastic time-dependent local fields~\cite{ernstSpinDiffusionSolids1998, grasserUnderstandingDynamicsRandomly2023}:
\begin{equation*}
H_{1}(t) = I^{z}_{i}B_{i}(t) + I^{z}_{j}B_{j}(t)
\end{equation*}
where
\begin{equation*}
B_{i}(t) = \sum_{k\neq i,j}d_{ik}I^{z}_{k}(t)
\end{equation*}
and similarly for $B_{j}(t)$. Transforming into the interaction frame of $H_{1}(t)$, $H_{ij}$ becomes 
\begin{equation*}
\tilde{H}_{ij} = -\frac{1}{4}d_{ij}[I^{+}_{i}I^{-}_{j}e^{i\phi_{ij}(t)} + I^{-}_{i}I^{+}_{j}e^{-i\phi_{ij}(t)}]
\end{equation*}
where
\begin{equation*}
\phi_{ij}(t) = \int_{0}^{t} \Omega_{ij}(t') dt'
\qquad \text{and} \qquad
\Omega_{ij}(t) = B_{i}(t) - B_{j}(t).
\end{equation*}
Applying second order perturbation theory and assuming the noise is stationary and Gaussian, then in the long-time limit the transition rate is given by:
\begin{equation*}
W_{ij} = \frac{d_{ij}^{2}}{8\Gamma_{ij}}.
\end{equation*}
where 
\begin{equation*}
\Gamma_{ij} = \int_{0}^{\infty} \langle \Omega_{ij}(\tau)\Omega_{ij}(0) \rangle d\tau.
\end{equation*}
Ignoring correlations between bath spins gives 
\begin{equation*}
\Gamma_{ij} = g_{n}\sum_{k\neq i,j} (d_{ik}-d_{jk})^{2}
\end{equation*}
where 
\begin{equation*}
g_{n} = \int_{0}^{\infty}\langle I^{z}(t)I^{z}(0) \rangle dt.
\end{equation*}
Assuming exponential decay of the bath correlation function gives $g_{n} = \tau_{n}/4$ where $\tau_{n}$ is the bath correlation time. Hence, 
\begin{equation*}
W_{ij} = \frac{d_{ij}^{2}}{2\tau_{n}\sum_{k\neq i,j} (d_{ik}-d_{jk})^{2}}
\end{equation*} 
$\tau_{n}$ is estimated from the second moment of the dipolar coupling distribution~\cite{abragamPrinciplesNuclearMagnetism1983}:
\begin{equation*}
\tau_{n} = \sqrt{\frac{\pi}{8}}\frac{1}{\sqrt{M_{2}}}
\end{equation*}
where
\begin{equation*}
M_{2}=\Big\langle \sum_{j\neq i} d_{ij}^{2}\Big\rangle_i,
\end{equation*}
which gives $\tau_n \approx 6.67$ ms. Using the expression above, the prefactor $\mathcal{A}$ in Eq.~\ref{eq:dipolar_rate} in the main text is given by
\begin{equation*}
\mathcal{A} = \frac{1}{2\tau_{n}\sum_{k\neq i,j}(d_{ik}-d_{jk})^{2}}.
\end{equation*}
Averaging this expression over the disordered lattice yields a bare hopping prefactor of $\mathcal{A}\approx 1 \times 10^{-4}$. Under the Floquet spin-locking sequence used in the experiments of Ref.~\cite{selcoEmergentDecoherenceDynamics2025a}, the effective dipolar Hamiltonian is reduced by a factor of two, resulting in a factor of four reduction in the hopping rates and an effective hopping prefactor of $\mathcal{A}\approx 2.5 \times 10^{-5}$, which is the value used throughout this work. This value is also consistent with the hopping prefactor previously extracted by fitting the experimental relaxation data in Ref.~\cite{selcoEmergentDecoherenceDynamics2025a}.

\subsection*{Appendix B: Coarse graining of strongly coupled spin clusters}
In this appendix, we examine whether the coherent intra-cluster dynamics discussed in Sec.~III~D affect the long-time transport behavior described by the continuous-time random walk (CTRW) model. As discussed in the main text, strongly coupled clusters support rapid coherent polarization exchange, whereas the CTRW description represents these dynamics by an effective incoherent hopping process. Since the anomalous transport reported in this work is governed by the much slower escape of polarization between clusters, one expects that a coarse-grained description in which each rapidly equilibrating cluster is treated as a single effective state should recover the same long-time dynamics. Such a description also reduces the size of the hopping network, providing a computationally simpler model for studying long-time transport. To test these ideas, we construct such a coarse-grained model and compare its transport properties with those of the original CTRW.

To construct the coarse-grained model, the strongly coupled clusters must first be identified. We use the same thresholding procedure introduced in Sec. IV of the main text. Given a hopping matrix $\mathbf{W}$, we choose a hopping-rate threshold $W_{\rm thresh}$ and define the binary connectivity matrix
\begin{equation*}
\begin{aligned}
B_{ij}(W_{\rm thresh})=\begin{cases}
    1, & W_{ij}>W_{\rm thresh}, \\
    0, & W_{ij}\leq W_{\rm thresh},
\end{cases}
\end{aligned}
\end{equation*}
where connected components of the resulting graph define the clusters.

Unlike the percolation analysis of Sec. IV, the goal here is not to identify the kinetic percolation threshold. Instead, we seek a threshold for which the dynamics within each cluster are much faster than the dynamics governing escape between clusters, thereby justifying the coarse-grained description. For a given value of $W_{\rm thresh}$, we first identify the clusters and then analyze two characteristic timescales. First, we define the cluster escape time, $\tau_{\rm escape}$, as the time spent within a cluster before the random walker exits to a different cluster. This quantity is obtained directly from CTRW trajectories by recording the residence time between entering and leaving a given cluster. Additionally, we define the intra-cluster equilibration time, $\tau_{\rm eq}$, as the characteristic timescale over which polarization equilibrates within a cluster. To estimate this quantity, we first construct the intra-cluster hopping matrix by restricting $\mathbf{W}$ to the sites belonging to a given cluster. We then compute the eigenvalues of this submatrix and define $\tau_{\rm eq}$ as the inverse magnitude of the smallest nonzero eigenvalue, corresponding to the slowest relaxation mode within the cluster. Averaging over disorder realizations and trajectories, we then examine the ratio $\langle \tau_{\rm escape}/\tau_{\rm eq} \rangle_{W_{\rm thresh}}$ as a function of $W_{\rm thresh}$. We note that, in strongly coupled clusters, the true polarization exchange rate is set by coherent dynamics which occur at a rate $d_{ij} > W_{ij}=\mathcal{A}d_{ij}^{2}$. Thus, the $\tau_{\rm eq}$ obtained from $\mathbf{W}$ is an upper bound estimate of the true intra-cluster equilibration time.

We find that $W_{\rm thresh}=1$ provides an averaged ratio $\langle \tau_{\rm escape}/\tau_{\rm eq} \rangle_{W_{\rm thresh}} \approx 50$, indicating that polarization equilibrates rapidly within clusters before escaping. This separation of timescales provides a natural criterion for coarse graining and motivates the choice of $W_{\rm thresh}=1$ used throughout the remainder of this appendix. We note that this is not the only possible procedure for partitioning the network into effective clusters, and more sophisticated community detection and coarse-graining approaches can provide a more systematic partitioning of the network~\cite{girvanCommunityStructureSocial2002, newmanModularityCommunityStructure2006, gfellerSpectralCoarseGraining2007, delvenneStabilityGraphCommunities2010}. However, the present procedure is sufficient for the purpose of this appendix and a more detailed investigation of optimal coarse-graining strategies is left for future work. To provide a physical reference scale for the chosen threshold, we note that $W_{\rm thresh}=1$, together with Eqs.~\ref{eq:dipolar_rate} and \ref{eq:dipolar_coupling} and neglecting the angular dependence of the dipolar interaction, corresponds to a separation that includes the first six nearest-neighbor shells of a site within the diamond lattice. Since first nearest-neighbor hopping is forbidden by the diamond crystal orientation ($3\cos^2\theta-1=0$), the resulting coarse-grained clusters typically range from one to five spins.

Having defined the strongly coupled clusters, we next construct a coarse-grained hopping matrix, $\mathbf{W}^{\rm CG}$. We assume that polarization equilibrates rapidly within each cluster prior to escape, so that for a given cluster $A$, the polarization is uniformly distributed over its $N_{A}$ sites. Under this assumption, the effective transition rate from cluster $A$ to cluster $B$ is given by
\begin{equation*}
\begin{aligned}
(\mathbf{W}^{\rm CG})_{BA} = \frac{1}{N_{A}}\sum_{i \in A}\sum_{j \in B}W_{ij},
\end{aligned}    
\end{equation*}
where $W_{ij}$ is the microscopic hopping rate between spins $i$ and $j$. Thus, the escape rate from a cluster is simply the average, over all sites within the originating cluster, of the microscopic hopping rate into the destination cluster. The diagonal elements are chosen to conserve probability,
\begin{equation*}
\begin{aligned}
(\mathbf{W}^{\rm CG})_{AA} = -\sum_{B \neq A}(\mathbf{W}^{\rm CG})_{BA}.
\end{aligned}    
\end{equation*}
We then perform the same CTRW simulations described in the main text using the coarse-grained rate matrix $\mathbf{W}^{\rm CG}$.

\begin{figure}[t]
    \centering
    \includegraphics[width=\columnwidth]{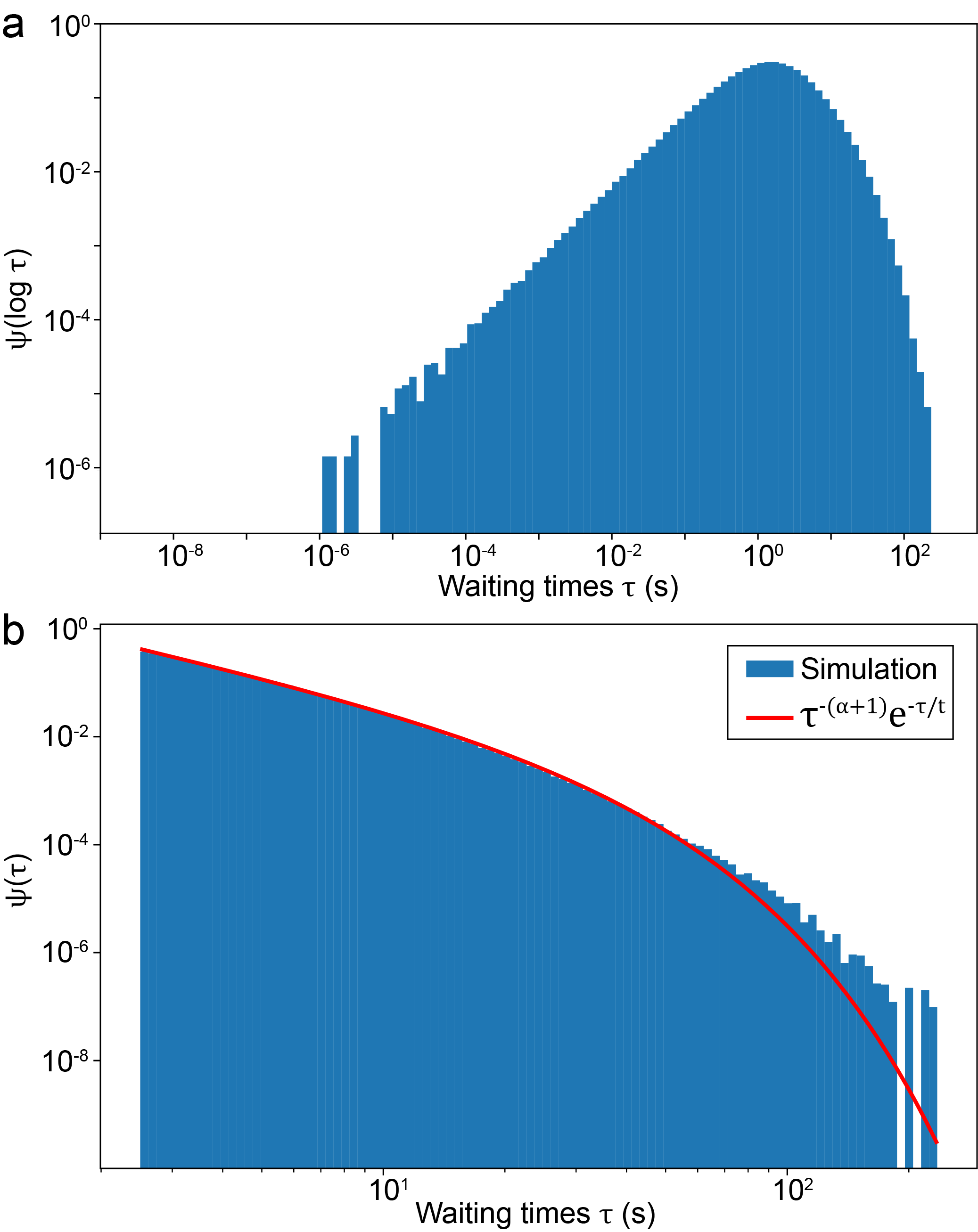}
    \caption{(a) Waiting-time distribution plotted as $\psi(\log \tau)$ obtained from the coarse-grained hopping matrix $\mathbf{W}^{\rm CG}$. The very short waiting-time events present in the microscopic description (Fig.~\ref{Fig2}a of the main text) are absent. (b) $\psi(\tau)$ for $\tau>2.5$~s; red line shows fit to Eq.~\ref{eq:waiting_tail}, yielding $\alpha=0.66$ and $t_{\rm cutoff}=17$ s, in close agreement with the values obtained from the microscopic description in the main text. Simulations were performed using the same parameters as Fig.~\ref{Fig2}(a,b) in the main text; $\approx1100$ spins per realization, $10^{3}$ trajectories per disorder realization, and $10^{3}$ independent disorder realizations.}
    \label{appendix_fig1}
\end{figure}

The effect of the coarse-graining procedure on the waiting-time distribution is shown in Fig.~\ref{appendix_fig1}. Panel (a) shows the waiting-time distribution, plotted as $\psi(\rm log\tau)$, analogous to Fig.~\ref{Fig2}a in the main text. As expected, the very short waiting-time events present in the microscopic CTRW description are absent. These short-time events originate from strongly coupled intra-cluster hops, where the underlying dynamics are better described by coherent dipolar dynamics rather than an incoherent Poisson process. In the coarse-grained description, these rapid intra-cluster dynamics are effectively averaged out, leaving only the slower escape dynamics between clusters, which are well described by an effective Poisson process.

The removal of these short-time processes does not significantly affect the long-time behavior of the waiting-time distribution. In panel (b), we show a magnified view of the tail of the distribution $\psi(\tau)$ and fit the tail to Eq.~\ref{eq:waiting_tail}, analogous to Fig.~\ref{Fig2}b in the main text. The resulting parameters, $\alpha=0.66$ and $t_{\rm cutoff}=17$ s, are in close agreement with the corresponding values obtained from the full microscopic CTRW model in the main text, $\alpha=0.64$ and $t_{\rm cutoff}=19$ s. This demonstrates that the long-time distribution of escape events, which controls the anomalous transport behavior, is insensitive to the precise microscopic description of the dynamics within strongly coupled clusters.

\begin{figure}[b]
    \centering
    \includegraphics[width=\columnwidth]{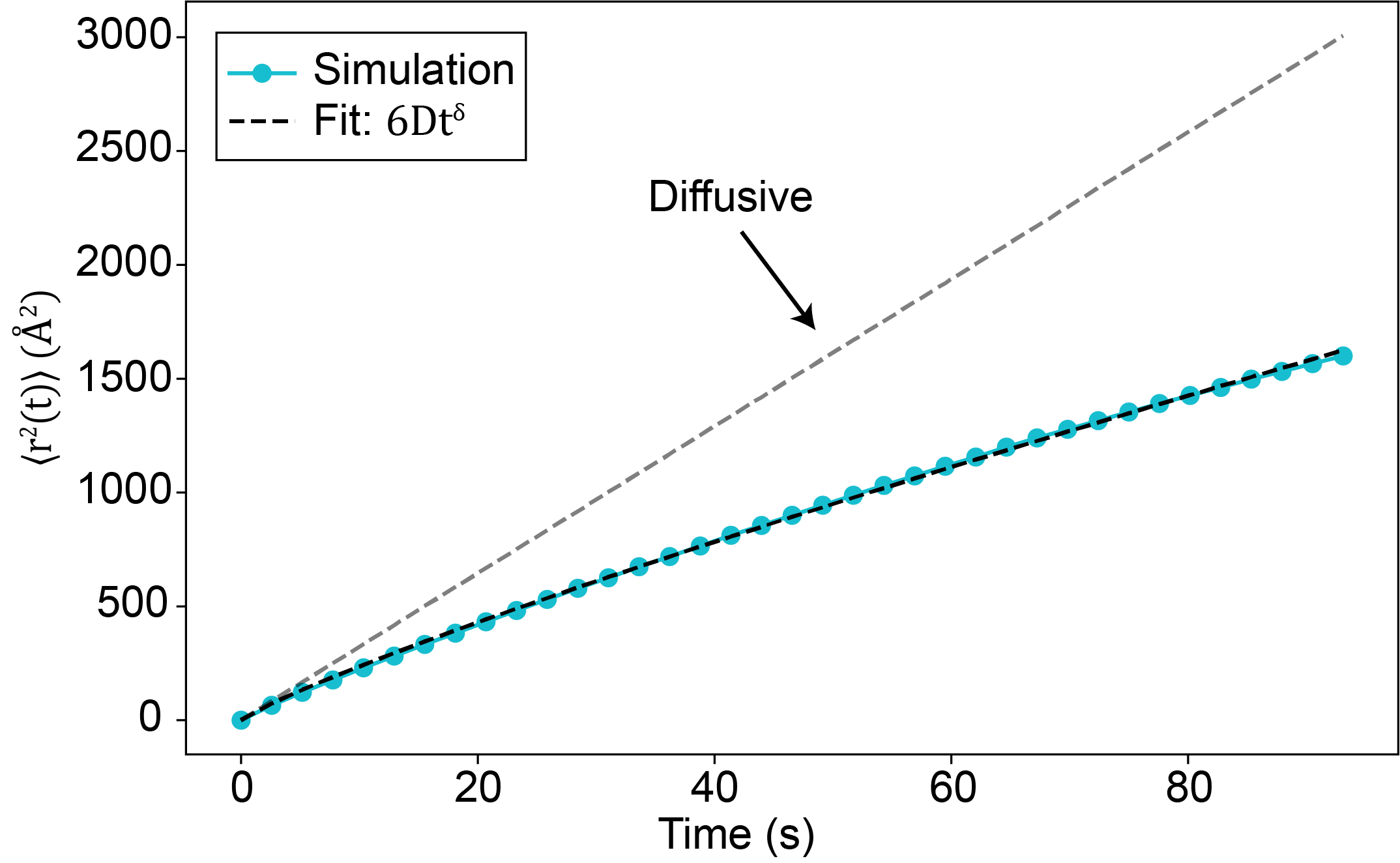}
    \caption{MSD as a function of time obtained using the coarse-grained hopping matrix $\mathbf{W}_{\rm CG}$. The black dashed line shows a fit to Eq.~\ref{eq:msd_time}, yielding $D_{\delta}=5.38~\mathrm{\AA}^{2}\mathrm{s}^{-\delta}$ and $\delta=0.86$, in agreement with the values obtained from the full microscopic model in the main text. Gray dashed line shows linear-in-time scaling expected for Fickian diffusion. Simulations were performed using the same parameters as Fig.~\ref{Fig3}(a,b) in the main text; $\approx3200$ spins per realization, $10^{3}$ trajectories per disorder realization, and $10^{2}$ independent disorder realizations.}
    \label{appendix_fig2}
\end{figure}

We further compare the transport behavior of the coarse-grained model through the MSD. To assign spatial coordinates to the coarse-grained states, each cluster is represented by the center-of-mass position of the spins comprising that cluster (computed with periodic boundary conditions), and the MSD is calculated using these effective cluster positions. Fig.~\ref{appendix_fig2} shows the MSD as a function of time, analogous to Fig.~\ref{Fig3}b in the main text. Fitting the MSD to Eq.~\ref{eq:msd_time} gives $D_{\delta}=5.38~\mathrm{\AA}^{2}\mathrm{s}^{-\delta}$ and $\delta=0.86$, in close agreement with the corresponding values obtained from the full microscopic CTRW model in the main text, $D_{\delta}=5.22~\mathrm{\AA}^{2}\mathrm{s}^{-\delta}$ and $\delta=0.87$. Thus, the long-time subdiffusive behavior is preserved after coarse-graining out the intra-cluster dynamics.

The MSD scaling with respect to the number of steps provides additional insight into the effect of the coarse-graining procedure. For the coarse-grained model, we obtain $\gamma=0.78$, compared with $\gamma=0.56$ for the full microscopic CTRW model. This difference is expected because the coarse-grained description removes the numerous short-distance hops occurring within strongly coupled clusters. In the limit where all rapidly equilibrating intra-cluster motion is coarse-grained into effective states, the distinction between step number and physical time becomes reduced, and the step-based scaling exponent is expected to approach the time-based exponent.

Together, these results demonstrate that the anomalous transport behavior identified in the main text is not a consequence of the microscopic description of the strongly coupled clusters. Instead, it arises from the broad distribution of inter-cluster escape times, which remains unchanged under the coarse-graining procedure. The coarse-grained model captures this transport while averaging over rapid intra-cluster equilibration, thus avoiding the need to explicitly describe the strongly coupled intra-cluster dynamics and reducing computational complexity.


\bibliography{RandomWalks}  

\end{document}